\documentclass[aip,jcp,floatfix,reprint]{revtex4-1}
\newcommand{\dSum}{\displaystyle\sum}
\newcommand{\dProd}{\displaystyle\prod}
\newcommand*{\plimsoll}{{\ensuremath{-\kern-4pt{\ominus}\kern-4pt-}}}
\usepackage{graphicx}
\usepackage{amsmath}
\usepackage{amssymb}
\usepackage{amsfonts}
\usepackage{color}
\begin{document}
\preprint{APS/123-QED}
\title{Inferring bulk self-assembly properties from simulations of small systems with multiple constituent species and small systems in the grand canonical ensemble}
\author{Thomas E. Ouldridge}
\affiliation{$^1$Rudolf Peierls Centre for Theoretical Physics, 1 Keble
Road, Oxford, OX1 3NP, UK }
\date{\today}
\begin{abstract}
In this paper we generalize a methodology [T. E. Ouldridge, A. A. Louis, and J. P. K. Doye, J. Phys.: Condens. Matter {\bf 22}, 104102 (2010)] for dealing with the inference of bulk properties from small simulations of self-assembling systems of characteristic finite size. In particular, schemes for extrapolating the results of simulations of a single self-assembling object to the bulk limit are established in three cases: for assembly involving multiple particle species, for systems with one species localized in space and for simulations in the grand canonical ensemble. Furthermore, methodologies are introduced for evaluating the accuracy of these extrapolations. Example systems demonstrate that differences in cluster concentrations between simulations of a single self-assembling structure and bulk studies of the same model under identical conditions can be large, and that convergence on bulk results as system size is increased can be slow and non-trivial. 
\end{abstract}
\maketitle
\section{Introduction}
\label{intro}
Self-assembly of monomer units into clusters of characteristic finite size is a central theme of biological and soft-matter systems. Examples include the formation of spherical micelles, \cite{Moroi1992, Gelbart1994} the self-assembly of virus capsids,\cite{FraenkelConrat55, Parent05, Zandi04, Zlotnick03, Zlotnick05, Zlotnick1999, Zlotnick2000, Casini04, Lavelle2009} the hybridization of DNA \cite{Santalucia2004, Owczarzy2004, Markham2008} and the formation of protein complexes.\cite{Jones2_1995, Goodsell2000, Levy2006} With increased computing power and improved simulation techniques, it has become possible to simulate mesoscale models that reproduce such self-assembling behaviour. In recent years mesoscopic models have been used to assemble micelles,\cite{Viduna98, Gottberg1997, Milchev2001, Zehl2006, Bolhuis1997, Kusaka2001, Pool2005, tenWolde1998, Kim2001, Floriano99, Panagiotopoulos02, Sciortino2009, Verde2010, Marrink2000} vesicles, \cite{Noguchi2001, Marrink2003, Sciortino2009} hollow shells of specific symmetry analogous to virus capsids\cite{Hagan2006,Wilber2007,Wilber09,Rapaport2008,Nguyen2007,Wilber09b,Johnston2010,Williamson2011,Mahalik2012} and aggregates of particles which resemble protein clusters.\cite{Villar09} Additionally, reflecting the growth of DNA nanotechnology,\cite{Pinheiro2011} many coarse-grained models of DNA assembly have recently been proposed.\cite{Jayaraman2007, Ouldridge2009, Ouldridge_tweezers_2010, Ouldridge2011,Sulc2012, Sambriski2008, Sambriski2009, Prytkova2010, Araque2011, Freeman2011, Tito2010, Allen2011, Hoefert2011, Schmitt2011, Linak2012} 

In many cases, these simulations report quantitative measures of assembly, such as the fraction of particles involved in clusters of a certain size. Experiments and thermodynamic theories typically involve bulk systems with a large number of particles, capable of forming many target structures. So, ideally, simulations should be performed on large systems from which the concentrations of various cluster sizes can be directly extracted. In many cases, however, it is not practical to simulate a large system, particularly when a large free-energy barrier suppresses equilibration. Such free-energy barriers often arise when monomers interact in a complex fashion (such as in DNA self-assembly), or when many monomers must cooperatively interact to create a stable target structure. Techniques such as umbrella sampling \cite{Torrie1977} allow systems to equilibrate despite these barriers, but they are not suited to biasing the formation of a large number of targets. 

As a consequence, it is common practice to simulate the assembly of a single target structure and attempt to infer bulk properties from the small-system data.\cite{Kusaka1998, Bolhuis1997, Kusaka2001, Pool2005, tenWolde1998, Wilber2007, Wilber09, Wilber09b, Ouldridge2009, Ouldridge_tweezers_2010, Ouldridge2011, Sambriski2008, Sambriski2009, Prytkova2010, Araque2011, Freeman2011, Tito2010, Allen2011, Hoefert2011, Schmitt2011, Romano2012} One obvious drawback of this approach is that any interactions between clusters (or tendencies to aggregate into macroscopic objects) are not observed. In many cases, however, monomers are in dilute solution and interactions between correctly formed targets are largely short-ranged and repulsive, and so these effects may be negligible. 
\begin{figure}
\begin{center}
\includegraphics[width=6cm]{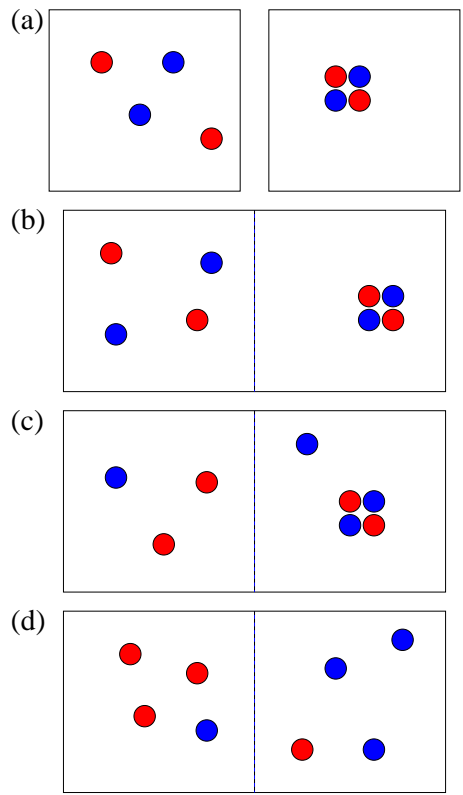}
\caption{\footnotesize{Illustration of the neglected concentration fluctuations that lead to finite size effects in canonical simulations. (a) Two examples of states sampled during the assembly of a single tetramer from two distinct monomer types. Now consider the assembly of two tetramers in twice the volume. For the purposes of analysis, separate this this larger cell into two halves, as in (b), (c) and (d). Some configurations of the system will have two particles of each type on either side of the partition, as in (b). These configurations will give the same average concentration of clusters as the original system of a single tetramer, as the available states in each half of the system are equivalent to the available states in (a). By contrast, configurations such as those in (c) and (d), in which the particles are unevenly distributed either side of the partition, will have a different average concentrations of clusters. This difference arises because the available states on either side of the partition are not captured by the single-target system.}}
\label{cluster fluctuations}
\end{center}
\end{figure}

If inter-target interactions are indeed negligible, can the average density (yield) of clusters in a small simulation be taken directly as the yield in a bulk system of the same total concentration? Not if simulations are performed in the canonical ensemble. Although the average density of particles is the same as in bulk, local fluctuations are not captured. For example, figure \ref{cluster fluctuations} demonstrates that the statistics of an eight-particle system in volume $2v$ are not accurately captured by those of a four-particle system in volume $v$. 

\begin{figure}[t]
\begin{center}
\includegraphics[width=6cm, angle=-90]{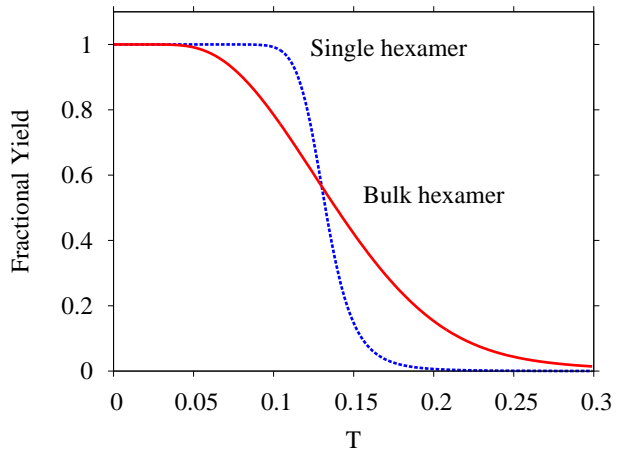}
\caption{\footnotesize{Melting transition for a cooperative toy model of hexamer formation from six identical monomers. The dashed curve is the fractional yield as a function of temperature for a single hexamer formed from six particles in the canonical ensemble. This curve represents a toy two-state model with a ratio of hexamer to monomer states given by $\Phi= \exp(20 - 0.4 T)$. The solid curve is the same transition extrapolated to the bulk limit.}}
\label{cooperative_transition_1}
\end{center}
\end{figure}

To emphasize the scale of the errors that can arise from assuming that the yield of clusters in a single-target simulation corresponds to the yield that would be measured for the same model in bulk, we consider a toy example of cooperative hexamer formation. Let us assume we are simulating a system of six particles in the canonical ensemble, and that these particles are found as either a hexamer or six isolated monomers, with the ratio of hexamer states to monomer states given by $\Phi = \exp(20 - 0.4 T)$, with $T$ as the temperature -- this model then has simple two-state behaviour. Note that this model is cooperative in the sense that the formation of a single hexamer is a cooperative phenomenon, requiring all 6 particles to be present, rather than that the presence of hexamers favours the formation of other hexamers. The yield curve of this small system as a function of temperature is plotted in Figure \ref{cooperative_transition_1}. 

We can then ask the question: what happens if we simulate a much larger number of monomers, in a volume such that the average density is the same as in the small simulation? The result actually depends on whether the hexamer consists of six identical particles or contains a number of distinct species. In the first case, which was treated in Reference \onlinecite{Ouldridge_bulk_2010}, the bulk yield of the model can be inferred assuming separate clusters behave ideally, and results are plotted in Figure \ref{cooperative_transition_1}. An outline of this calculation is also presented in Section \ref{Inference in CE}. It is clear that, for this toy model, the bulk yield is very different from the small-system yield, the biggest difference being that the transition is far wider in bulk than in a single-target system. In other words, conditions that generate high or low yields of clusters in a single-target system tend to generate less extreme yields in the bulk limit. Clearly, if one wishes to compare simulation data of a model to bulk experiment, it is important to estimate bulk yields of the model correctly before doing so. We note that if the assumption of ideality of separate clusters is not reasonable, the exact form of the deviation between small- and large-system yields will vary from that presented here: nonetheless, the cause of the discrepancy will persist and differences will remain large.

In Reference \onlinecite{Ouldridge_bulk_2010}, we used a statistical mechanical approach to derive the correct extrapolation procedure from small simulations of clusters formed from identical monomers, and dimers formed by distinct particles. The convergence on bulk yields as simulation size increased was also studied, enabling the approximations inherent in the extrapolation to be checked for any particular system. In this work we extend the methodology to include arbitrary size clusters with any number of particle types, including cases in which one of the constituent monomers is immobilized.

Grand canonical simulations, which in principle incorporate these local concentration fluctuations, also give misleading results if only a single large cluster is sampled. The size of these errors and a methodology for correcting them are also presented here, and compared to an alternative approach common in the literature.

The techniques for extrapolating from small simulations presented in this work can be seen as methodologies for estimating free energies of assembly for a model system, from which equilibrium concentrations follow. This work emphasizes the substantial errors that can arise when simulation yields are not properly analysed, derives a methodology for performing this analysis and presents an approach for examining the accuracy of the assumption of ideality underlying the theory for a given system. Additionally, we show how simulation yields converge on bulk values as the system size is increased, allowing an estimate of how large a simulation must be before finite-size effects are negligible.

We emphasize that the methods discussed here will not necessarily make any given model agree better with experimental data: rather, they allow an accurate and physically meaningful comparison with experiment to be made. Failure to accurately account for finite size effects when matching a simulation of a small system to bulk experiments or thermodynamic theories is analogous to using a flawed algorithm to obtain the data. The results obtained are not a true representation of the bulk behaviour of the model being simulated. 

The rest of this paper is structured as follows. Firstly, the assumptions and formalism relevant to this study are introduced in Section \ref{definitions}. Next, finite size corrections for arbitrarily sized clusters of multiple particle species are derived and the convergence on bulk yields analysed in Section \ref{Finite size effects in the Canonical Ensemble}. The case of systems in which one of the reactants is localized within a certain volume is then considered in Section \ref{localization}. Subsequently, issues arising in grand-canonical simulations are discussed in Section \ref{Gcanonical}, and finally the inference of bulk properties other than cluster yields is outlined in Section \ref{other properties}.
\subsection{Methodology, assumptions and definitions}
\label{definitions}
This paper uses a statistical mechanical approach to highlight differences between simulations of assembly of a single target cluster and simulations in which multiple targets can form, both in the canonical and grand canonical ensembles, and introduces an approach for relating them. The terminology and assumptions of this analysis are introduced here. To demonstrate the statistical finite-size corrections, several model systems are simulated in this work -- as the details of each simulation are different, and because the simulations are only for illustrative purposes, the methodology is briefly described at the relevant point in the text and covered in detail in the supplementary material.\cite{supplementary} 

We consider simulations in a small periodic cell of volume $v$ of interacting objects which are generically referred to as {\em particles}. We assume that the particles have some tendency to aggregate into clusters of finite size (rather than undergo a thermodynamic phase transition). It is assumed that, for the system in question, we have some way of defining which particles are in a {\em cluster}. The details of such a definition are not important -- we only require that clustering involves particles being in close proximity, and strongly bound particles will be part of the same cluster. It is also assumed that, in a bulk system at the appropriate concentration, clusters behave ideally ({\it i.e.,} interactions between clusters are negligible except when they form larger clusters). We define a macrostate to consist of all states with a given set of clusters, whether in a small or large system. We now define some quantities that will be of use in the analysis. The notation is more complex than in Reference \onlinecite{Ouldridge_bulk_2010}, but it is also more powerful, allowing systems with an arbitrary number of particle species to be analysed. 
\begin{itemize}
\item $D$ is an (arbitrary) scale factor relating a thermodynamically large volume to a simulation volume $v$.
\item $y$ is the number of distinct monomer species present.
\item ${\{i\}} = {(i_1,i_2, ..., i_y)}$ defines a cluster containing $i_j$ particles of type $j$. On several occasions, it will be necessary to take a sum or product over all clusters. In these cases, $\{m\} = (m_1,m_2, ..., m_y)$ will be used as a set of dummy variables. For cluster formation from a single species, $\{i\}$ reduces to a single integer $(i)$.
\item $\eta_{\{i\}}$ is the number of times that the cluster $\{i\}$ appears in a macrostate. The macrostate is therefore completely specified by the set $\{\eta\}$. Let $\eta_j$ be an abbreviation for the number of isolated monomers of particle-type $j$.
\item $z_{\{i\}}$ is the partition function for cluster $\{i\}$ in the simulation volume $v$, with the internal degrees of freedom treated {\em distinguishably}. The need to use expressions involving both distinguishable and indistinguishable statistics arises in this work because the enumeration of configurations is most easily done by treating particles distinguishably, and then accounting for indistinguishability afterwards. Furthermore, computer simulations naturally treat objects as distinguishable, which is particularly important in grand canonical simulations. In this work, partition functions calculated with distinguishable statistics will be symbolized by $z$ or $Z$, and indistinguishable partition functions represented by $q$ or $Q$.
\item $Z_{\{\eta\}}$ is the partition function of a system of volume $v$ when in a macrostate $\{\eta\}$. This partition function is calculated using {\em distinguishable statistics}.
\item $q_{\{i\}}$ is the partition function for a cluster ${\{i\}}$ in a volume $Dv$, with the internal degrees of freedom treated {\em indistinguishably}. $q_j$ is an abbreviation for the partition function of a monomer of particle-type $j$ in volume $Dv$.
\item $[{\{i\}}]$ is the concentration of cluster ${\{i\}}$, and $[j]$ is the concentration of a monomer of particle type $j$. Here it is most convenient to use particles per unit volume as the measure of concentration.
\item $\mu_{\{i\}}$ is the chemical potential of cluster ${\{i\}}$. Let $\mu_j$ be an abbreviation for the chemical potential of an isolated monomer of particle-type $j$.
\end{itemize}

\section{Systems in the canonical ensemble}
\label{Finite size effects in the Canonical Ensemble}
\subsection{Multi-species assembly}
A number of authors have taken cluster yields in canonical simulations of single-target assembly as directly applicable to bulk systems.\cite{Wilber2007, Wilber09b, Sambriski2008, Sambriski2009, Prytkova2010, Freeman2011} These examples involve dimers and clusters of a single species, which we analysed in our earlier work.\cite{Ouldridge_bulk_2010} Interesting structures, however, are not exclusively dimers or formed from identical subunits. DNA nanostructures, such as polyhedra,\cite{Goodman2005, He2008} often involve several different single strands. Virus capsids can require more than one type of coat protein, and some work has been undertaken to simulate models of such structures.\cite{Johnston2010} Simulators have also considered templated assembly, in which distinct shells of particles form cooperatively. \cite{Williamson2011} Here we extend our previous work to monodisperse assemblies of an arbitrary number of different species, so that quantitative analyses of such systems can be performed using data from small simulations. For small simulations of monodisperse assembly in the canonical ensemble, it is natural to use {\em exactly} enough particles to form a single target. Our discussion and examples will assume this is the case, although the results do not depend on it.

\subsection{Bulk equilibrium yields of self-assembly}
\label{Bulk equilibrium yields of self-assembly}
In this section we derive expressions for bulk cluster yields in terms of the quantities defined in Section \ref{definitions} that will later be used to analyse small simulations. The results can be slightly simplified for assemblies of a single species: these simplifications are highlighted in the text. A standard result of equilibrium statistical mechanics of ideal particles is that the chemical potential $\mu_{\{i\}}$ of cluster-type ${\{i\}}$ in a volume $Dv$ is given by
\begin{equation}
\mu_{{\{i\}}} = -{\rm k_B}T \frac{\partial}{\partial \eta_{\{i\}}} \ln \left ( \frac{q_{\{i\}}^{\eta_{\{i\}}}}{\eta_{\{i\}}!} \right ) \approx -{\rm k_B}T \ln \left ( \frac{q_{\{i\}}}{\eta_{\{i\}} }\right ),
\label{mu_i}
\end{equation}
where the approximation becomes an equality in the thermodynamic limit.\cite{Huang1987} 
In this limit, equilibrium thermodynamics gives $\sum_{\{i\}} \nu_{\{i\}} \mu_{\{i\}} = 0$ for any possible reaction,\cite{Huang1987} where $\nu_{\{i\}}$ are the stoichiometric coefficients of the species in the process. For the special case of a cluster $\{i\}$ forming from its constituent monomers, the relation between the $\mu_{\{i\}}$ reduces to $\sum_j i_j \mu_j = \mu_{\{i\}}$. Combining this result with Equation \ref{mu_i}, we obtain
\begin{equation}
\frac{ \eta_{\{i\}}}{\prod_j^y \eta_j^{i_j}} = \frac{q_{\{i\}}}{\prod_{j}^y q_{j}^{i_j}}
\label{thermo_multi_particle}
\end{equation}
for the equilibrium between a cluster $\{i\}$ and its constituent monomers. We convert to concentration using the system volume $Dv$, giving
\begin{equation}
\frac{ [{\{i\}}]}{\prod_j^y [j]^{i_j}} = (Dv)^{i_{\rm tot}-1} \frac{q_{\{i\}}}{\prod_{j}^y q_{j}^{i_j}} = v^{i_{\rm tot}-1} \psi_{\{i\}},
\label{thermo_multi_particle_2}
\end{equation}
where $i_{\rm tot} = \sum_x i_x$, and we have defined $\psi_{\{i\}}$ for later convenience. In the case of cluster formation from only one species of particle, the products in the denominators contain only one term. The quantities $v^{i_{\rm tot}-1} \psi_{\{i\}}$ are model properties (related to binding strengths), and are independent of our small simulation volume $v$: $\psi_{\{i\}}$, which can be extracted from small simulations, have a $v$ dependence that cancels with the explicit $v^{i_{\rm tot}-1}$. 

If $v^{i_{\rm tot}-1} \psi_{\{i\}}$ are known, it is possible to extract the equilibrium yields of all clusters by solving the simultaneous equations given by Equation \ref{thermo_multi_particle_2} (one for each cluster of more than one particle) and the conservation of total particle number, 
\begin{equation}
\sum_{\{m\}} m_x [{\{m\}}] = [x]_{\rm T},
\label{conc_yield}
\end{equation}
where $[x]_{\rm T}$ is the total concentration of particles of type $x$. One conservation equation is obtained for each distinct species of particle. This result holds for any set of initial concentrations $[x]_{\rm T}$. Further, it is even possible to infer bulk yields in the isothermal-isobaric ensemble, provided our assumptions of ideality remain valid. In the isothermal-isobaric system at pressure $p$, the total volume is not fixed. Equations \ref{thermo_multi_particle_2} and \ref{conc_yield} still hold due to the equivalence of ensembles in the thermodynamic limit, and they are solved along with 
\begin{equation}
p/k_{\rm B}T = \sum_{\{m\}} [\{m\}],
\label{pressure}
\end{equation}
which follows from the equation of state of ideal gases and allows the total volume as well as the concentrations to be determined. Note that the temperature $T$ at which results are inferred must be the same as that used to determine $v^{i_{\rm tot}-1} \psi_{\{i\}}$, which will generally be $T$-dependent in a non-trivial manner. In either ensemble, the problem of obtaining bulk yields reduces to obtaining the quantities $v^{i_{\rm tot}-1} \psi_{\{i\}}$ and then solving a set of simultaneous equations. We note that, as $v^{i_{\rm tot}-1} \psi_{\{i\}}$ are constants for a given $T$, Equation \ref{thermo_multi_particle_2} is a classic `law of mass action' as expected for a simple assembling system.\cite{Huang1987} 

\subsection{Appropriate ensembles and free energies}
\label{ensembles}
Experiments are often performed under conditions of approximately constant particle number, temperature and pressure. This would suggest that the use of the isothermal-isobaric ensemble is appropriate, and indeed this is true for assembly processes studied in the gas phase, such as in Reference \onlinecite{Kusaka1998}. In this case, it would be more natural to convert the concentrations into partial pressures $p_x = k_{\rm B}T[x]$. If a standard pressure $p^\plimsoll$ is introduced, we can convert Equation \ref{thermo_multi_particle_2} into
\begin{equation}
\frac{ p_{\{i\}}/p^\plimsoll }{\prod_j^y (p_{j}/p^\plimsoll)^{i_j}} = \psi_{\{i\}} \left(\frac{vp^\plimsoll}{kT}\right)^{i_{\rm tot}-1},
\end{equation}
where $ p_{\{i\}}$ is the partial pressure of cluster ${\{i\}}$. This is a well-known result (the law of mass action for an ideal system) and allows us to define a temperature-dependent dimensionless equilibrium constant $K^\plimsoll(T) $, with an associated free energy change of formation $\Delta G^\plimsoll_{\{i\}}$:
\begin{equation}
\psi_{\{i\}} \left(\frac{vp^\plimsoll}{kT}\right)^{i_{\rm tot}-1} = K^\plimsoll (T)= \exp(-\Delta G^\plimsoll_{\{i\}}/RT).
\end{equation}
Measuring the quantities $v^{i_{\rm tot}-1} \psi_{\{i\}}$ is therefore equivalent to finding the standard free energy change of formation at a given temperature.

In many cases of experimental interest, however, the assembling particles are not isolated: other species are present, contributing to the total pressure of the system. If the interaction of these extra species with the self-assembling species is negligible, and their partial pressure is known, Equation \ref{pressure} can simply be modified to $({p-p^\prime})/k_{\rm B}T = \sum_{\{m\}} [\{m\}]$, in which $p^\prime$ is the partial pressure of the non-reactant species. 

Many examples of self-assembly, including the majority of soft matter systems mentioned in Section \ref{intro}, occur in dilute solution. A dilute solution is a case in which the partial pressure of the solvent dominates that of the self-assembling particles, and in general the interactions of the solvent with the self-assembling particles are non-negligible. Many of the mesoscopic models discussed in Section \ref{intro} treat the solvent implicitly. \cite{Viduna98,Gottberg1997,Milchev2001,Zehl2006,Verde2010,Noguchi2001,Wilber2007,Wilber09,Rapaport2008,Nguyen2007,Wilber09b,Johnston2010,Williamson2011,Jayaraman2007, Ouldridge2009, Ouldridge_tweezers_2010, Ouldridge2011, Sambriski2008, Sambriski2009, Prytkova2010, Araque2011, Freeman2011, Tito2010, Allen2011, Hoefert2011, Schmitt2011, Linak2012} 
With the solvent treated implicitly, model clusters (at low enough concentrations) will behave ideally except for when they bind to form a larger cluster. A self-consistent methodology for comparing these models to experiment would be to assume that clustering causes no change to the pressure of the system: {\em i.e.,} the partial pressure of solute is negligible and that the change of the solvent/solute interaction due to clustering has no effect on pressure. In this case, clustering does not influence the system volume and so the cluster yields in bulk are the same for the canonical and isothermal-isobaric ensemble. Yields can then be inferred using a fixed volume, requiring only Equations \ref{thermo_multi_particle_2} and \ref{conc_yield}. For dilute solutions, it is common to use concentrations rather than partial pressures: introducing a standard concentration $[c]^\plimsoll$, the standard free energy change of formation $\Delta G^\plimsoll_{\{i\}}$ follows from Equation \ref{thermo_multi_particle_2} as
\begin{equation}
(v[c]^\plimsoll])^{i_{\rm tot}-1} \psi_{\{i\}} = K^\plimsoll = \exp(-\Delta G^\plimsoll_{\{i\}}/RT).
\end{equation}
Note that although relative volume changes due to cluster formation in real systems may be small, meaning that the assumption of a constant total volume is reasonable, the $pV$ (pressure-volume) contribution to the Gibbs free energy of cluster formation may not be negligible. In this case, mesoscale models with implicit solvents that are compared to experimental data would still neglect any volume change, but would incorporate the $pV$ contribution to assembly implicitly as part of the effective interaction between particles.

Some approaches, including fully atomistic representations, explicitly model solvent particles.\cite{Bolhuis1997,Kusaka2001,Pool2005,tenWolde1998,Kim2001,Floriano99,Panagiotopoulos02,Marrink2000,Marrink2003,Rapaport2008} Simulations of such models can be analysed in terms of the solute clustering, treating the solvent implicitly at the level of the analysis rather than in the actual model. Single-target simulations performed at constant volume will neglect any $pV$ contributions to assembly inherent in the model -- with this caveat, bulk yields can be estimated through the methodology presented in this work. The cluster yields of single-target simulations of explicit solvent models in isobaric ensembles will also require statistical finite-size corrections. If the variation in volume is negligible compared to the overall volume, then the methodology presented in this work can be used to estimate bulk yields by taking the average volume in simulations as the small-system volume. 

For the remainder of this work, the analysis will be presented in terms of dilute solutions (as this is most relevant to our work), and hence the bulk yield in the canonical ensemble is the appropriate quantity for comparison with experiment. Nonetheless, obtaining $\psi_{\{i\}}$ allows the calculation of isobaric yields if desired. To avoid the complication of multiple $(v[c]^\plimsoll)^{i_{\rm tot}-1}$ factors, it is simpler to analyse the problem in terms of the quantities $q_{\{i\}}$ and $\psi_{\{i\}}$ and convert to molar concentrations afterwards. The majority of this work is focused on correctly estimating $\psi_{\{i\}}$ from single-target simulations. 
\subsection{Inferring bulk yields from small canonical simulations}
\label{Inference in CE}
To extract $\psi_{\{i\}}$ from small canonical simulations, it is helpful to calculate the contribution to the small-system partition function of a macrostate with $n_j$ particles of type $j$, arranged into the set of clusters $\{\eta\}$. Under our assumptions, this is
\begin{equation}
Z_{\{\eta\}} = \dProd_{\{m\}} \frac{ \left(z_{\{m\} } \right)^{\eta_{\{m\}}}}{\left( \eta_{\{m\}}\right)! \left( \Pi_x^y m_x!\right)^{\eta_{\{m\}}} } {\dProd_j^y n_j!}.
\label{v_sim_1}
\end{equation}
This expression is obtained by multiplying the individual distinguishable partition functions $z_{\{m\} }$ together, then considering all possible permutations of identical particles which change the clustering. Dividing by ${\prod_j^y n_j!}$ would make the statistics indistinguishable. For a single constituent species, the products over $x$ and $j$ contain only one term. The yield of a certain cluster $\{i\}$ in the small simulation, $v [{\{i\}}]_{(1)}$, follows from $Z_{\{\eta\}}$ and is given by
\begin{equation}
v [{\{i\}}]_{(1)}= \frac{\sum_{\{ \eta\}} \eta_{\{i\}} Z_{\{\eta\}}}{\sum_{\{ \eta\}} Z_{\{\eta\}}}.
\label{v_sim_2}
\end{equation} 
Here the sum runs over all possible macrostates $\{ \eta\}$: the denominator is then the entire partition function of the $n$-particle system. The subscript in $v [{\{i\}}]_{(1)}$ indicates that we are considering a single-target system. Multiplying the concentration by the original simulation volume $v$ means that yields reported are the average numbers of different clusters in a volume $v$, a convenient dimensionless quantity.

One can relate the $q_{\{i\}}$ to $z_{\{i\}}$ by incorporating the relative scale factor $D$ of the volumes in which they are defined, and accounting for the over-counting of indistinguishable states within $z_{\{i\}}$. We obtain
\begin{equation}
\frac{q_{\{i\}}}{D} = \frac{z_{\{i\}}}{\prod_x^y i_x !}.
\label{qz_multi_particle}
\end{equation}
In the case of cluster formation from a single type of particle, the product over $x$ contains only one term.
Combining Equations \ref{thermo_multi_particle_2}, \ref{v_sim_1}, \ref{v_sim_2} and \ref{qz_multi_particle} then yields
\begin{equation}
v[{\{i\}}]_{(1)}= \frac{ \dSum_{\{ \eta\}} \eta_{\{i\}}
\dProd_{\{m\}} \frac{ \left(\psi_{\{m\} } \right)^{\eta_{\{m\}}}}{\left( \eta_{\{m\}}\right)! }} 
{\dSum_{\{ \eta\}} 
\dProd_{\{m\}} \frac{ \left(\psi_{\{m\} } \right)^{\eta_{\{m\}}}}{\left( \eta_{\{m\}}\right)! }}. 
\label{D_equals_1_equation}
\end{equation}
We have therefore expressed the small system yield as a function of the ratios $\psi_{\{m\}}$, or equivalently $\psi_{\{i\}}$ (as $m$ and $i$ are just labels), which determine the bulk yield. $\psi_{\{i\}}$ can therefore be extracted from a small simulation by fitting the observed yields $v[{\{i\}}]_{(1)}$ to Equation \ref{D_equals_1_equation}, and the bulk yields obtained as discussed in Section \ref{Bulk equilibrium yields of self-assembly}. For the case of homoclusters (clusters consisting of one species of particle), an alternative method that does not require fitting and automatically decouples the simultaneous equations is possible, as outlined in Reference \onlinecite{Ouldridge_bulk_2010}.
\begin{figure}[t]
\begin{center}
\includegraphics[width=6cm, angle=-90]{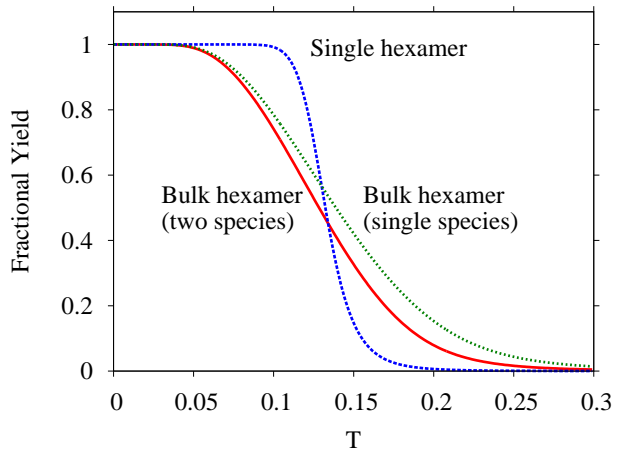}
\caption{\footnotesize{Completely cooperative transition for a hexamer formed from six monomers. The blue curve is the yield as a function of temperature for a single hexamer in the canonical ensemble following a two-state model with a ratio of hexamer to monomer states given by $\Phi= \exp(20 - 0.4 T)$. The green curve is the same transition extrapolated to the bulk limit (with the same total concentration of particles) in the case where the hexamer is formed from identical particles, and the red curve is the extrapolation to bulk for the same small system result when the hexamer is formed from two different species, contributing three particles each. }}
\label{cooperative_transition_2}
\end{center}
\end{figure}

It is instructive to reconsider the toy model of cooperative hexamer formation introduced in Section \ref{intro}. The results of Figure \ref{cooperative_transition_1} follow directly from Equations \ref{thermo_multi_particle_2}, \ref{conc_yield} and \ref{D_equals_1_equation}. The ratio of hexamer to monomer states in a small simulation is given by $\Phi= \exp(20 - 0.4 T)$, with $T$ as the temperature.
\begin{itemize}
\item $v[(6)]_{(1)} = \Phi / (1+ \Phi)$, $v[(1)]_{(1)} = 6 / (1+ \Phi)$.
\item $\psi_{(6)} = \Phi/6! $ is found by substituting the single-target yields of the toy model into Equation \ref{D_equals_1_equation} ($\psi_{(1)}=1$ by definition). Note that in this case Equation \ref{D_equals_1_equation} can be solved for $\psi_{\{i\}}$ (rather than requiring fitting) due to the simplicity of the system, but this is not generally the case.
\item The bulk fraction of hexamers, $f=v[(6)]$, can then be shown to obey $6! f = 6^6 \Phi (1-f)^6$ using $\psi_{(6)}$ and Equations \ref{thermo_multi_particle_2} and \ref{conc_yield}.
\item This equation can be solved numerically to give a bulk fractional yield of hexamers $f=v[(6)]$ that can be compared to the single-target yield $v[(6)]_{(1)}$.
\end{itemize}

Alternatively, we can imagine a {\em different} system in which the hexamer consists of three particles each of two different species. Let us again assume that in single-target simulations the system follows a two-state model with a ratio of hexamer to monomer states given by $\Phi= \exp(20 - 0.4 T)$. 
\begin{itemize}
\item $v[(3,3)]_{(1)} = \Phi / (1+ \Phi)$, $v[(1,0)]_{(1)} = v[(0,1)]_{(1)} = 3 / (1+ \Phi)$.
\item $\psi_{(3,3)} = \Phi/(3!)^2 $ is found by substituting the single-target yields of the toy model into Equation \ref{D_equals_1_equation} ($\psi_{(1,0)} = \psi_{(0,1)}=1$ by definition). Once again, Equation \ref{D_equals_1_equation} can be solved for $\psi_{\{i\}}$ in this simple case, rather than requiring a fit.
\item The bulk fraction of hexamers, $f=v[(3,3)]$, can then be shown to obey $(3!)^2 f = 3^6 \Phi (1-f)^6$ using $\psi_{(3,3)}$ and Equations \ref{thermo_multi_particle_2} and \ref{conc_yield}.
\item This equation can be solved numerically to give a bulk fractional yield of hexamers $f=v[(3,3)]$ that can be compared to the single-target yield $v[(3,3)]_{(1)}$.
\end{itemize}

The extrapolations for these two different systems (with the same single-target yield) are plotted in Figure \ref{cooperative_transition_2}, along with the single-target yield.
As with clusters of one particle type, the bulk transition is far broader than in the single-target case. This widening effect can be understood by considering the effect of concentration fluctuations, as illustrated in Figure \ref{cluster fluctuations}. If a system of twice the size is considered, fluctuations in concentration like that in Figure \ref{cooperative_transition_1}\,(c) can occur. Such fluctuations strongly favour the formation of exactly one target structure rather than zero or two, as it is impossible to form one on the left of the box and on the right hand side, the extra particle makes the formation of a cluster much more statistically favourable. Increasing the system size and allowing concentration fluctuations within cells of volume $v$ therefore tends to give yields that are less dominated by one particular cluster size than the single-target system. The result is a much broader transition in bulk.

It is also possible to understand why the heterocluster yield is lower in bulk than for homoclusters with the same single-target yield. For heteroclusters in bulk, we have fluctuations of concentration in a volume $v$ not only of the total particle number, but also of the relative number of each type of particle, as shown in Figure \ref{cooperative_transition_1}\,(d). These fluctuations always disfavour the formation of target clusters, resulting in a lower yield in bulk for the same single-target yield. 

Although we do not claim that the methodology presented in this work will make a given model agree better with an experiment, it is worth noting that single-target yields do not obey the law of mass action as would be expected for simple models, unlike bulk yields. As a simple example, consider a dimer-forming system of two distinct particles. The law of mass action, as embodied by Equation \ref{thermo_multi_particle_2}, predicts that $[(1,1)] \propto [(1,0)][(0,1)]$. For a stoichiometric solution, this equation reduces to $[(1,1)] \propto [(1,0)]^2$. In a single-target system, $[(1,1)]_{(1)}/[(1,0)]_{(1)} \propto 1/v$ as doubling the volume with the same number of particles will halve the ratio of bound to unbound states. However, $1/v = [(1,0)]_{\rm T}$, the total concentration of particle of type 1. Therefore, $[(1,1)]_{(1)} \propto [(1,0)]_{(1)}[(1,0)]_{\rm T}$, which is a fundamentally different result from the law of mass action.

\subsection{Convergence on bulk yields}
\label{convergence}
It is instructive to consider how yields converge on their bulk values as system size is increased (whilst maintaing the same total concentration of particles). Firstly, this gives an idea of how large simulations must be to reflect the thermodynamic limit. Secondly, it provides a tool to check the validity of the extrapolation in Section \ref{Inference in CE}: if it is possible to simulate the formation of two targets in twice the volume, the change in yield from the first simulation can be compared to the predictions of this section to ensure that the assumptions underlying the theory are accurate.

Consider a simulation of a system of size $d$ with the same total concentration as the relevant single-target system, where $d$ is not necessarily thermodynamically large. We can extend the concepts of the previous section, in which an expression for the yield of clusters for $d=1$ was found in terms of $\psi_{\{i\}}$, in a very simple fashion to give
\begin{equation}
v[{\{i\}}]_{(d)}= \frac{ \dSum_{\{ \eta\}} \eta_{\{i\}}
\dProd_{\{m\}} \frac{ \left(\psi_{\{m\}} / d^{m_{\rm tot} -1} \right)^{\eta_{\{m\}}}}{\left( \eta_{\{m\}}\right)! }} 
{d \dSum_{\{ \eta\}} 
\dProd_{\{m\}} \frac{ \left(\psi_{\{m\}}/ d^{m_{\rm tot} -1} \right)^{\eta_{\{m\}}}}{\left( \eta_{\{m\}}\right)! }}.
\label{D_greater_1_equation}
\end{equation}
Here the division by powers of $d$ corrects for the larger volume. An alternative useful quantity is the fraction of particles of type $a$ that are found in clusters of type $\{i\}$ as a function of simulation size $d$, $f^a_{\{i\}(d)} = \frac{i_a}{n_a}v[{\{i\}}]_{(d)}$, where $n_a$ is the number of particles of type $a$ in the single-target simulation.
For a given set of $\psi_{\{i\}}$, one can explicitly calculate $f^a_{\{i\} (d)}$ and observe its convergence on bulk values. In all systems we have studied, $f^a_{ \{i\}_(d)} - f^a_{ \{i\} (\infty)}$ scales as $1/d$ at sufficiently large $d$, although convergence at low $d$ can be more complex.
\begin{figure}
\begin{center}
\includegraphics[width=6.5cm, angle=-90]{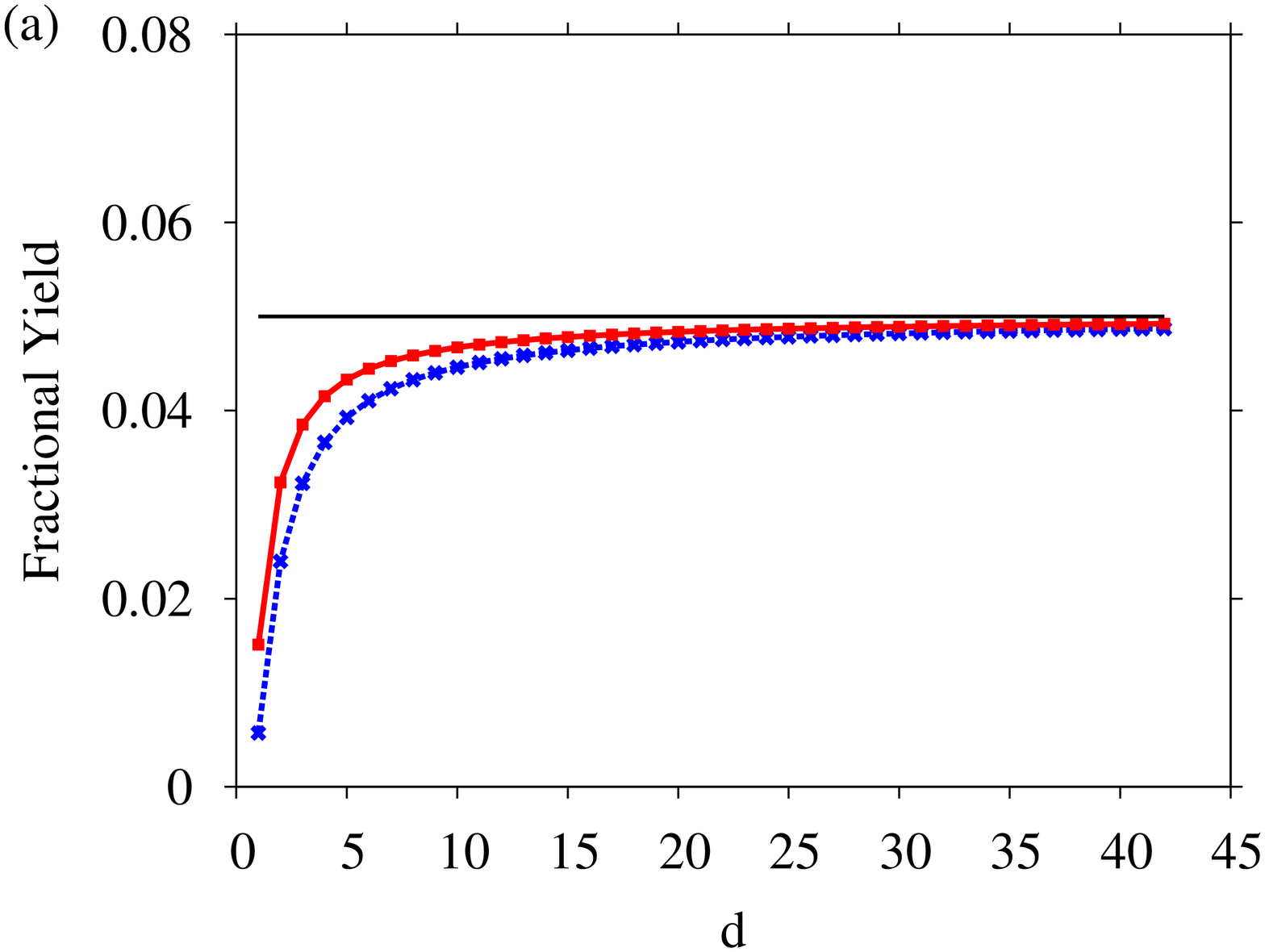}\\
\includegraphics[width=6.5cm, angle=-90]{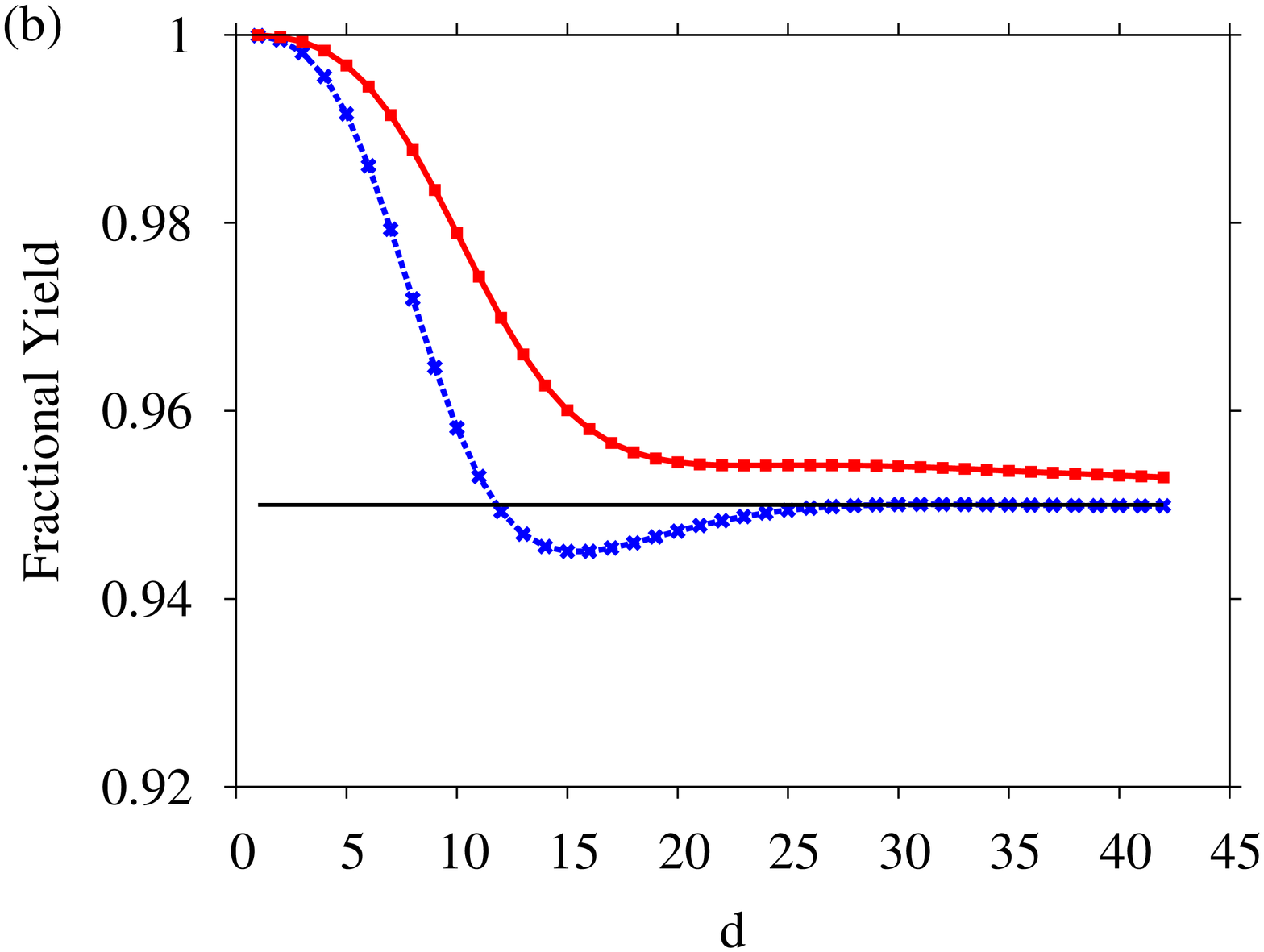}
\caption{\footnotesize{ Convergence on bulk yields for cooperative hexamer formation as a function of system size $d$. (a) low hexamer yield (5\,\% in bulk). (b) high hexamer yield (95 \,\% in bulk). The solid curve depicts convergence for hexamers consisting of two distinct species, each contributing three particles to the hexamer (in this case $\psi_{(3,3)} = 9.33 \times 10^{-5}$ and $ 8.34 \times 10^4 $ for the 5\,\% and 95 \,\% cases respectively). The dashed curve shows the result for hexamers containing six identical monomers (with $\psi_{(6)} = 1.30 \times 10^3$ and $1.46 \times 10^{-6}$ for the 5\,\% and 95 \,\% cases respectively ). }}
\label{heterocluster_conv}
\end{center}
\end{figure}

To make more concrete statements, consider the completely cooperative toy model of hexamer formation introduced in Section \ref{intro}. For a given bulk yield, $\psi_{(3,3)}$ (or $\psi_{(6)}$ in the single-species case) can be inferred from Equations \ref{thermo_multi_particle_2} and \ref{conc_yield}, and then substituted into Equation \ref{D_greater_1_equation} to give the fractional yield as a function of system size. Even in this simple case, two qualitatively distinct regimes of convergence are observed, at high and low yield of the target structure, as shown in Figure \ref{heterocluster_conv}. At low yield, convergence is monotonic and quickly settles down to the $1/d$ form. By contrast, convergence at high yield is initially slow, before a more rapid decay towards the bulk value. In some cases, converge can involve oscillations before the $1/d$ regime is reached.

These results are qualitatively comparable to equivalent size homoclusters. It is noticeable that, for the same high bulk yield of target clusters, heterocluster convergence is slower and has less pronounced oscillations than for homoclusters. Convergence is slower because, in order to generate the same high target yield in the infinite limit, the single-target simulation must have a higher ratio of target clusters to monomers ($\Phi$) for heteroclusters than homoclusters (as can be seen in Figure \ref{cooperative_transition_2}, and was discussed in Section \ref{Bulk equilibrium yields of self-assembly}), and convergence to the bulk value is then slower. By contrast, heterocluster convergence is slightly better at low target yields, as this time the need to have a higher $\Phi$ to obtain the same bulk yield reduces the error.

The oscillations at high yield for homoclusters coincide with the points at which macrostates with a certain number of target clusters come to dominate the ensemble. Initially, the macrostate with $d$ target clusters (all particles are found in clusters of the largest size) is dominant. Eventually, as system size is increased, the macrostate with $d-1$ target clusters becomes dominant due to the entropic cost of having no monomers. 
However, the $d-1$ macrostate becomes dominant before $(d-1)/d = f^a_{\{t\} (\infty)}$ (the fractional yield of target clusters in the bulk limit), and consequently $f^a_{\{t\} (d)} < f^a_{\{t\} (\infty)}$. As $d$ increases further, the $d-1$ macrostate remains dominant but now $(d-1)/d > f^a_{\{t\} (\infty)}$, and so $f^a_{\{t\} (d)} > f^a_{\{t\}(\infty)}$. Smaller oscillations are then repeated as macrostates with $d-2$, $d-3$ {\it etc.} targets successively become dominant. Eventually the oscillations are overwhelmed by the overall $1/d$ convergence. The suppression of oscillations in heteroclusters can be understood in terms of their slower convergence -- as the configurations with more monomers take longer to become dominant for a given bulk yield, the tendency to underestimate the bulk yield is suppressed.

Of course, real systems are not perfectly cooperative, and the finite concentration of intermediate clusters has consequences for the convergence properties. Again, we cannot claim to have tested all possibilities but the effect of intermediate cluster sizes appears to be similar to the effect in homoclusters.\cite{Ouldridge_bulk_2010} The consequences for convergence are most pronounced when the prevalent intermediate clusters are close in size to the majority cluster, when convergence is generally slower than if the intermediates are absent. 
\subsection{Example extrapolation}
\label{Example extrapolation C}
To demonstrate the use of the extrapolation technique on a model system, we consider the formation of three-armed DNA trimers from distinct strands, using the coarse-grained DNA model of Reference \onlinecite{Ouldridge_thesis}. This model treats DNA as a string of rigid nucleotides with effective interactions to model chain connectivity, excluded volume, hydrogen bonding and base stacking. In this work we are not really concerned with how good an approximation the model is to reality. We are simply demonstrating that the extrapolation procedure can be applied to real simulation results, giving bulk statistics that could then, if desired, be sensibly compared to experiment. 
\begin{figure}
\begin{center}
\includegraphics[width=8.7cm]{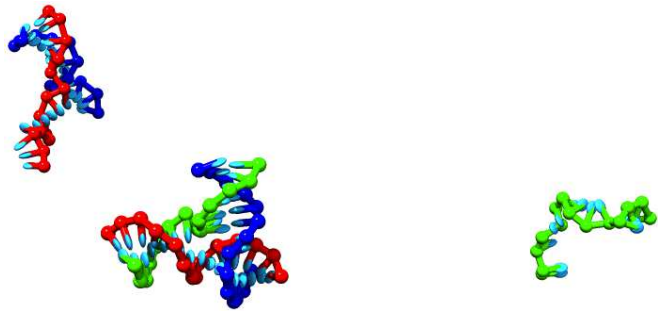}
\caption{\footnotesize{Snapshot from a simulation of a trimer-forming DNA system, using the DNA model of Reference \onlinecite{Ouldridge_thesis}. Backbones of strands of type $1$ are coloured red, $2$ are blue and $3$ are green; all bases are coloured sky blue. The cluster nearest to the centre is typical of a three-armed trimer, an isolated stand is shown on the far right and a two-strand intermediate is shown in the top left.}}
\label{trimer_figure}
\end{center}
\end{figure}
\begin{table}
\begin{center}
\begin{tabular}{c | c c| c c }
cluster $\{i\}$ & $v[{\{i\}}]^{\rm sim}_{(1)}$ & $v[{\{i\}}]^{\rm pred} $& $v[\{i\}]_{(2)}^{\rm sim}$ & $v[\{i\}]_{(2)}^{\rm pred}$ \\ 
\hline
(1,0,0) & 0.054(2) & 0.208(4) & 0.114(6) & 0.116(3)\\
(0,1,0) & 0.101(5) & 0.291(5) & 0.193(8) & 0.182(6)\\
(0,0,1) & 0.171(7) & 0.375(6) & 0.279(12) & 0.272(8)\\
(1,1,0) & 0.133(7) & 0.211(7) & 0.190(9)& 0.183(7)\\
(1,0,1) & 0.063(3) & 0.128(5) & 0.103(5) & 0.094(5)\\
(0,1,1) & 0.0157(7) & 0.0451(15) & 0.0295(17) & 0.0277(11)\\
(1,1,1) & 0.750(10) &0.4526(90) & 0.587(15)& 0.607(11) \\
\hline
\end{tabular}
\caption{\footnotesize Yields of DNA clusters in a trimer forming system. Clusters $\{i\}$ are defined by the number of each strand type that they contain: $(i_1,i_2,i_3)$ contains $i_j$ strands of type $j$. The definition of what constitutes a cluster is given in the supplementary material.\cite{supplementary} Yields of each cluster are shown for simulations of a single cluster ($v[{\{i\}}]_{(1)}^{\rm sim}$), and bulk results are extrapolated from these data using the methodology discussed in the text ($v[{\{i\}}]^{\rm pred}$). Yields from simulations of two clusters ($v[j]_{(2)}^{\rm sim}$) are compared to the yield $v[\{i\}]_{(2)}^{\rm pred}$, which is predicted from $v[{\{i\}}]_{(1)}^{\rm sim}$ using Equation \ref{D_greater_1_equation}.
\label{trimer_conv}}
\end{center}
\end{table}

We consider three distinct strands of DNA, the sequences of which are given in the supplementary material. \cite{supplementary} The three strands tend to form three-armed junctions, as each strand has two 6-base sections, each of which is complementary to a 6-base section on a different strand. An example of the three-armed junction is given in Figure \ref{trimer_figure}, which also shows a two-strand intermediate and an isolated strand.

We simulated one strand of each type in a periodic cell, measuring the resultant distribution of clusters. Details of the simulations are provided in the supplementary material.\cite{supplementary} The yields of various clusters resulting from these simulations are tabulated in Table \ref{trimer_conv}. Also shown are the yields predicted for bulk by extracting $\psi_{\{i\}}$ from fitting Equation \ref{D_equals_1_equation} to the data and then solving Equations \ref{thermo_multi_particle_2} and \ref{conc_yield} for $[\{i\}]$. As is evident, the bulk yields are significantly different from those in the single-target simulations: specifically, the high yield of trimers in the single-target simulation is reduced, and the lower yields of single strands and two-strand complexes are increased in bulk. This is as expected from the general broadening of the transition that was discussed in Section \ref{Inference in CE}, as the dominant cluster in a single-target simulation becomes less dominant in bulk.

This extrapolation assumes that separate clusters behave ideally. As the DNA model used here has only short-ranged interactions, and the system is fairly dilute, this seems a reasonable assumption. We can perform a more rigorous test, however, in that we have used the same methodology to predict not only the yield in the infinite system-size limit, but also how the yield changes as the system size is increased. The expected yield in a two-target simulation with the same total density, inferred using $\psi_{\{i\}}$ and Equation \ref{D_greater_1_equation}, is given in Table \ref{trimer_conv}. Although more challenging than the single-target simulation, it is also possible to simulate the simultaneous formation of two trimers in twice the volume using a high-dimensional reaction coordinate for umbrella sampling: additional information on these simulations is given in the supplementary material.\cite{supplementary} The resultant cluster yields are also shown in Table \ref{trimer_conv}.

As is evident from Table \ref{trimer_conv}, the predicted and measured yields in a two-target simulation are in excellent agreement. This strongly suggests that the assumptions of the extrapolation procedure (such as ideality) are reasonable for this model under these conditions, and therefore that the bulk values reported in Table \ref{trimer_conv} are representative of the yields that would follow from a macroscopically large simulation. Furthermore, it provides a `sanity check' of the accuracy of the approach presented in Sections \ref{Bulk equilibrium yields of self-assembly}, \ref{Inference in CE} and \ref{convergence}.

\subsection{Theory of localising a single reactant species}
\label{localization}
\begin{figure}
\begin{center}
\includegraphics[width=6cm]{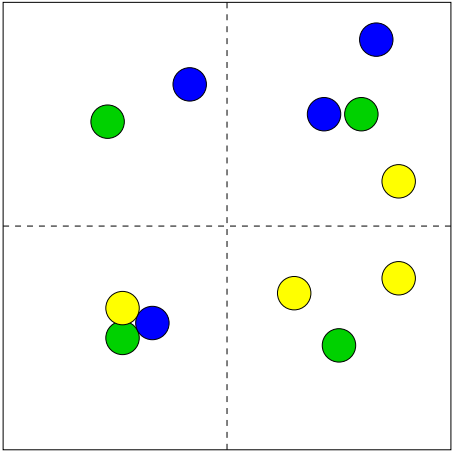}
\caption{\footnotesize{ Schematic depiction of a localized species. The green particles are localized near the centre of the cells, and hence their concentration does not fluctuate.}}
\label{tether figure}
\end{center}
\end{figure}

In some experimental systems, the particles that associate are not all free to diffuse. For example, DNA microarray assays consist of DNA `probes' which are tethered to a surface, and `target' molecules which diffuse through solution.\cite{Vasiliskov2001, Naiser2008} With the advent of DNA origami, experimentalists are now able to localize isolated reactants at will.\cite{Wickham2011} Figure \ref{tether figure} illustrates such a localisation for a generic system. Several groups have simulated the binding of DNA to a tethered strand, extracting quantitative estimates of melting temperatures without applying finite size corrections.\cite{Tito2010, Allen2011, Hoefert2011, Schmitt2011} We note that practical DNA microarrays typically have such a high density of strands tethered to the surface that clusters are unlikely to behave independently and ideally:\cite{Jayaraman2007} the simulations in References \onlinecite{Tito2010, Allen2011, Hoefert2011, Schmitt2011}, however, considered isolated tethered strands and hence can only be sensibly compared to much sparser systems. To perform this comparison, it is necessary to consider whether corrections must be applied to yields from single-target simulations. 

It is not {\it a priori} obvious whether tethering one reactant will change our earlier results, for which local concentration fluctuations were invoked to explain the difference between bulk and small-system statistics. Note that here we are not concerned with whether the mechanism of tethering interacts with the particles, either destabilizing or stabilizing the bound state. For example, the presence of a surface to which a particle is attached could be either attractive or repulsive for the non-localized particles. Instead, we are concerned with whether extrapolation to bulk for a given set of $z_{\{i\}}$ differs from Section \ref{Inference in CE}. 

To analyse this problem, it is instructive to consider how the standard result of $\sum_i \nu_i\mu_i=0$, with $\mu_i$ given by Equation \ref{mu_i} and $\nu_i$ being stoichiometric coefficients in a reaction, arises directly from the partition function. The contribution to the partition function (calculated using {\em indistinguishable} statistics) of a large system with a macrostate which has $\eta_{\{i\}}$ clusters of type $\{i\}$ (with all clusters behaving ideally) is given by
\begin{equation}
Q_{\{\eta\}}(D) = \dProd_{\{i\}} \frac{ q_{\{i\}}^{\eta_{\{i\}}}}{\eta_{\{i\}}!} = \dProd_{\{i\}} \frac{ (D z_{\{i\}}/\prod_x^y i_x!)^{\eta_{\{i\}}}}{\eta_{\{i\}}!}.
\end{equation}
This expression contains a product over all the partition functions of the individual clusters, divided by an $\eta_{\{i\}}!$ to avoid double counting of states, which must be included because each $q_{\{i\}}$ includes a separate integral for each cluster over the whole of the system volume. Maximizing $Q_{\{\eta\}} $ with respect to $\{\eta\}$ yields the standard result.

We now consider a system in which one of the reactants is immobilized (let this be particle type $1$, and let us further assume that only one particle of type 1 can be involved in any given assembly). The first consequence is that there is no need to divide by $\eta_{\{i\}}!$ when the cluster $\{i\}$ includes the immobilized species, as there is no tendency to count indistinguishable states twice when the clusters cannot move over all space. One must, however, still deal with combinatorial effects. In particular, we now have to calculate the combinatorial factor associated with the number of different choices of immobilized particles that are involved in cluster formation. This introduces a factor
\begin{equation}
\frac{n_{1}!}{\prod_{\{i\}, i_1 \neq 0} \eta_{\{i\}}!},
\end{equation}
in which $n_{1}$ is the total number of localized particles in the system.
The second effect is that $q_{\{i\}}$ does not scale with system volume for $i_1 \neq 0$, so that
\begin{equation}
q_{\{i\}} = \frac{z_{\{i\}}}{\prod_x^y i_x!} \hspace{5mm} { \rm for} \hspace{3mm} i_1 \neq 0.
\end{equation}
Including both alterations, we obtain the following partition function of a system with immobilized particles in the macrostate $\{N\}$, $Q^{\rm im.}_{\{N\}}$:
\begin{equation}
\frac{Q^{\rm im.}_{\{\eta\}}} {n_1 !} = \frac{ \dProd_{\{i\}, i_1=0} \frac{ q_{\{i\}}^{\eta_{\{i\}}}}{\eta_{\{i\}}!} \dProd_{\{i\}, i_1 \neq 0} { q_{\{i\}}^{\eta_{\{i\}}}} } {\dProd_{\{i\}, i_1 \neq 0} \eta_{\{i\}}!}.
\label{Q_im_1}
\end{equation}
In terms of $z_{\{i\}}$, Equation \ref{Q_im_1} becomes
\begin{equation}
\frac{Q^{\rm im.}_{\{\eta\}}}{n_1! } = \frac{ \dProd_{\{i\}, i_1=0} \left(\frac{D z_{\{i\}}}{\prod_x^y i_x!}\right)^{\eta_{\{i\}}} \dProd_{\{i\}, i_1 \neq 0} {\left( \frac{z_{\{i\}}}{\prod_x^y i_x!}\right)^{\eta_{\{i\}}}} } {\dProd_{\{i\}} \eta_{\{i\}}!}.
\end{equation}

It is trivial to check that for two sets of clusters $\{\eta\}$ and $\{\eta^\prime\}$, $Q^{\rm im.}_{\{\eta\}} / Q^{\rm im.}_{\{\eta^\prime\}}$ has exactly the same functional dependence on $z_{\{i\}}$, $\eta_{\{i\}}$ and $\eta_{\{i\}}^\prime$ as $Q_{\{\eta\}} / Q_{\{\eta^\prime\}}$. Therefore there are no statistical consequences of tethering one of the reactants -- a given set of $z_{\{i\}}$, which means a given set of yields in a single-target simulation, will extrapolate to bulk yields that are identical to the case in which all species are free to diffuse. In other words, the procedure of scaling single-target results to bulk is unchanged, although the single-target results themselves may be influenced by the localisation mechanism. We also note that at no stage have we used the fact that $D$ is thermodynamically large in this argument, so it applies just as well to systems of intermediate size. Therefore it is not only the scaling to the bulk limit that is unchanged by tethering, but also the form of the convergence on the bulk limit as system size is increased.

Practically, this means that Equations \ref{thermo_multi_particle_2} and \ref{conc_yield} can be directly used to calculate the expected bulk concentrations from any initial set of reactant concentrations, having used Equation \ref{D_equals_1_equation} to extract $\psi_{\{i\}}$ from a single-target simulation in a small volume. Note, however, that although the yields of clusters are expressed as concentrations, tethered clusters will not be uniformly distributed throughout the system. Similarly, the convergence of yields as the system size is increased at a constant total density can be followed using Equation \ref{D_greater_1_equation}. An example of such an extrapolation is provided in Section \ref{example localized}.

Physically, the result is identical to the unlocalized case because in this idealized limit the only important coordinates are the relative separations of cluster-forming particles. As a consequence, it is irrelevant that particles of type $1$ are tethered, as the concentration fluctuations of the other particles in the vicinity of type $1$ provide the same statistical correction as the untethered case. Such an argument does not hold if two particles that are involved in an assembly are localized. For a trivial counter-example, one could take heterodimer formation. Localizing both species will give thermodynamics identical to the small system limit. 

When performing simulations of this kind, it is possible that non-tethered particles will interact with the tethering mechanism in the unbound state. If this has a significant effect on the unbound partition function, it will lead to errors in the extrapolation. Changing the simulation volume and observing whether the statistics of bound states change in the expected way can check for such effects.

\subsection{Example extrapolation for a localized species}
\label{example localized}
\begin{figure}
\begin{center}
\includegraphics[width=8.8cm]{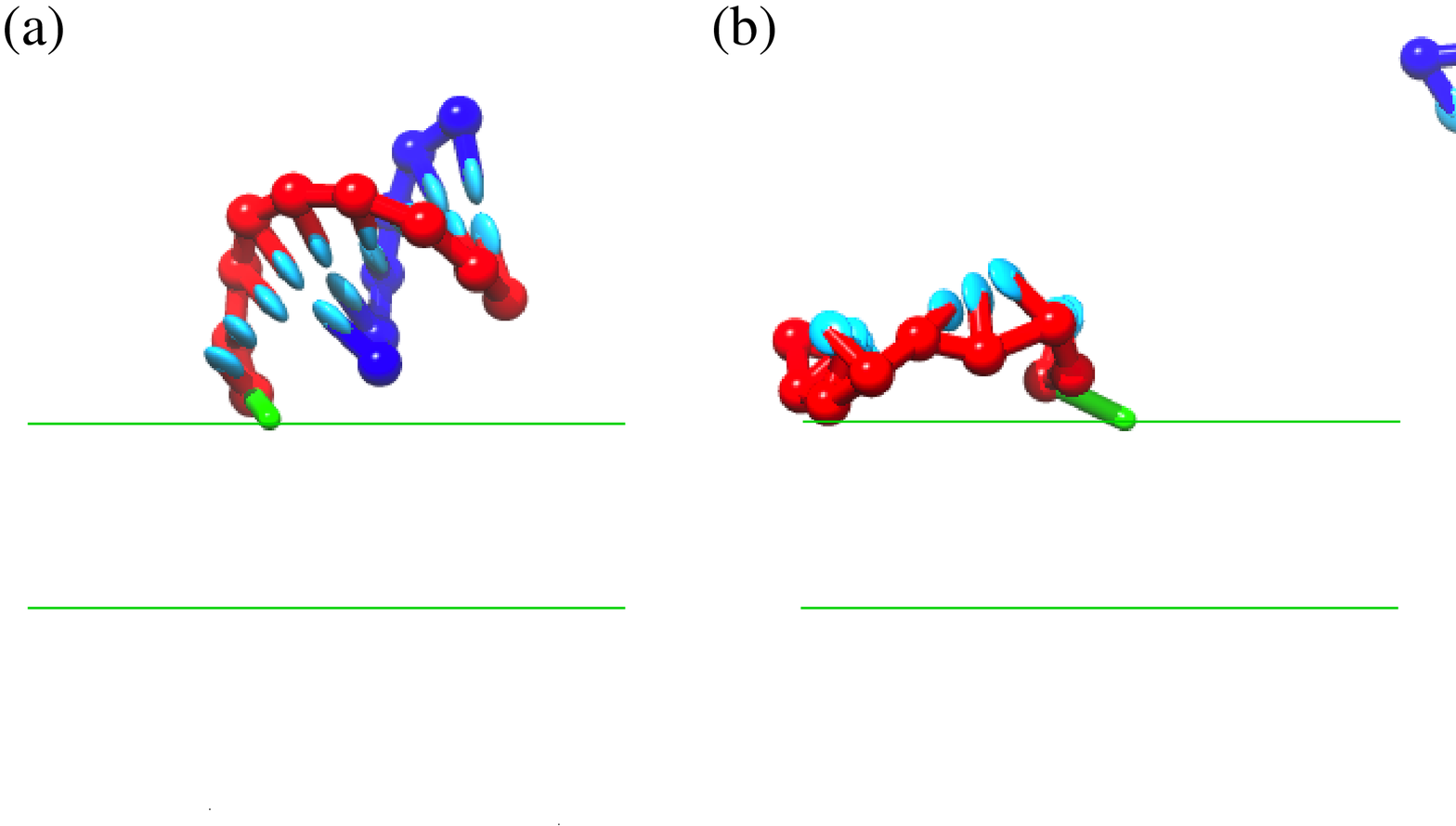}
\caption{\footnotesize{Two states from a simulation of DNA binding to a tethered strand. In both cases the longer red strand is attached to the surface by the $5^\prime$ end. The parallel lines represent the excluded volume of the surface: the centre of any backbone site is forbidden from entering this region. In (a), the two strands are bound; in (b), the shorter blue strand is detached and free to diffuse.}}
\label{tether figure 2}
\end{center}
\end{figure}

As a demonstration of extrapolation with a localized particle, we consider the formation of a six-base-pair DNA duplex using the model of Reference \onlinecite{Ouldridge_thesis}. One of the strands in this duplex has a three-base tail, which is permanently attached to a repulsive surface by its $5^{\prime}$ end (DNA strands are directional: the two ends are labelled $3^{\prime}$ and $5^{\prime}$). The other strand is free to diffuse. We performed simulations of a single-target system, and of a system of twice the volume and number of strands, using umbrella sampling to enhance equilibration. Figure \ref{tether figure 2} shows typical bound and unbound states from a single-target simulation, highlighting the tethering of the longer strand to a surface. Further details of the simulations are provided in the supplementary material.\cite{supplementary}

The yield of duplexes in a simulation volume, $v [\{i\}]_{(1)}^{\rm sim} =0.764(8)$, was obtained in the single-target simulations. This implies a ratio of $z_{\{1,1\}}/(z_{\{1,0\}}z_{\{0,1\}})=\psi_{\{1,1\}} =3.24(14)$, from which we infer an expected two-target yield of $v [\{i\}]_{(2)}^{\rm pred}= 0.667(8)$ using Equation \ref{D_greater_1_equation} and a bulk fraction of duplexes $v [\{i\}]^{\rm pred} = 0.578(7)$ using Equations \ref{thermo_multi_particle_2} and \ref{conc_yield}. Our measured two-target yield of dimers, $v [\{i\}]_{(2)}^{\rm sim}= 0.651(7)$, does not show a statistically significant difference from the prediction, suggesting that the bulk value is also reliable and that non-ideal effects are not detectable at this precision. As expected, the yield of the dominant cluster size (the duplex) is reduced due to the extrapolation, consistent with the general broadening effect of concentration fluctuations on transitions.

\section{Simulations in the grand canonical ensemble}
\label{Gcanonical}
\subsection{The technique}
As an alternative to canonical simulations, it is possible to use the grand canonical ensemble (at first, we consider a system with only one species of reactant). Instead of fixing the number of particles absolutely, one simulates a system in a volume $v$ such that configurations containing $n$ particles are sampled with the relative probability\cite{Frenkel2001}
\begin{equation}
P(n) \,\, \propto \,\, \frac{ {\rm e}^{\beta \mu n} Z(n)}{n!},
\end{equation}
where $Z(n)$ is the partition function of an $n$-particle system in volume $v$, calculated using {\em distinguishable} statistics, and the $n!$ accounts for distinguishability. $\mu$ is the chemical potential of the monomers, which regulates the average concentration. 

Restricting ourselves initially to one species of reactant, the set of numbers $\{j\}$, which identifies a cluster, is reduced to a single integer $(j)$, the cluster size. We retain the brackets for consistency with earlier notation. A macrostate $\{\eta\}$ is then observed in a simulation with probability
\begin{equation}
P(\{\eta\}) \propto \dProd_{j>0} \frac{z_{(j)}^{\eta_{(j)}}}{\eta_{(j)}! \,\, (j!)^{\eta_{(j)}}} {\rm e}^{\beta \mu j \eta_{(j)} },
\label{P_eta}
\end{equation}
where $z_{(j)}$ is the partition function of a cluster of size $j$ in the volume $v$, as before. The result follows from multiplying individual distinguishable partition functions $z_{(j)}$ together, including combinatorial factors to account for exchange of particles and multiplying by $ {\rm e}^{\beta \mu n }$, with $n$ being the total number of particles in $\{\eta\}$.

In principle, small grand canonical simulations are capable of capturing the concentration fluctuations highlighted in Section \ref{intro}. We will first demonstrate that the average concentrations found in a small volume in the grand canonical ensemble are identical to those in a bulk system with the same monomer chemical potential. The bulk equilibrium yields of clusters can be expressed in terms of the chemical potential of the monomers: Equation \ref{mu_i} implies
\begin{equation}
[(1)] = z_1 {\rm e}^{\beta \mu} / v,
\label{GC_1}
\end{equation}
and
\begin{equation}
[(j)] = [(1)]^j (Dv)^{j-1} \frac{q_{(j)}}{q_1^j} = v^{-1}\frac{z_{(j)}}{j!} {\rm e}^{\beta \mu j}
\label{v[k]}
\end{equation}
follows from combining Equations \ref{thermo_multi_particle_2}, \ref{qz_multi_particle} and \ref{GC_1}. We now consider the yield of clusters in a small grand canonical simulation of a volume $v$. Let $\bar{\eta}_{(j)}$ be the average number of clusters of type $(j)$ observed during simulation. As $P(\{\eta\})$ in Equation \ref{P_eta} factorizes into separate terms for each cluster type, the calculation of $\bar{\eta}_{(j)}$ does not involve the properties of clusters other than $(j)$. Therefore, using Equation \ref{P_eta}, we find
\begin{equation}
\bar{\eta}_{(j)}= \frac{\dSum_{\eta_{(j)}=0}^\infty \frac{\eta_{(j)} z_{(j)}^{\eta_{(j)} }\exp(\beta \mu j \eta_{(j)}) }{ \eta_{(j)}! (j!)^{\eta_{(j)}}} }{\dSum_{\eta_{(j)}=0}^\infty \frac{ z_{(j)}^{\eta_{(j)} }\exp(\beta \mu j \eta_{(j)}) }{ \eta_{(j)}! (j!)^{\eta_{(j)}}} } .
\label{eta_bar}
\end{equation}
This equation can be rewritten as 
\begin{equation}
\bar{\eta}_{(j)} =\frac{1}{\beta j} \frac{\partial }{\partial \mu} \ln \left( \dSum_{\eta_{(j)}=0}^\infty \frac{ \left( \frac{z_{(j)}}{j!} \exp(\beta \mu j) \right)^{ \eta_{(j)}}} { \eta_{(j)}!} \right).
\end{equation}
The sum inside the logarithm is actually the series expansion of an exponential, allowing the expression to be easily evaluated
\begin{equation}
\bar{\eta}_{(j)} =\frac{1}{\beta j} \frac{\partial }{\partial \mu} \left( \frac{z_{(j)}}{j!} \exp(\beta \mu j) \right) = \frac{z_{(j)}}{j!} {\rm e}^{\beta \mu j}.
\label{small_gc_yield}
\end{equation}
Dividing by $v$ to give a concentration shows that Equation \ref{small_gc_yield} is consistent with Equations \ref{GC_1} and \ref{v[k]}, and hence that a small grand canonical simulation should provide the same cluster concentrations as a bulk simulation of the same system. If desired, measured cluster concentrations can then be used to evaluate $\psi_{\{i\}}$ through Equation \ref{thermo_multi_particle_2}. With $\psi_{\{i\}}$, model systems can be compared to experiment under a range of conditions, as discussed in Section \ref{ensembles}. 

\subsection{Quantification of errors}
Cluster yields can only be estimated accurately, however, if multiple large clusters can exist simultaneously during the simulation. Here we define `large' to mean any cluster containing more than one particle. When assembly is difficult and biased sampling techniques are employed, it is often impractical to bias the formation of multiple large clusters. Consequently, states with multiple large clusters are never sampled and errors are introduced.
Here we derive how the observed concentration of clusters differs from the true yield if only a single large cluster is sampled. 

Let us assume that a simulation samples states that contain at most a single cluster of more than one particle. In this case, the average number of clusters of size $j > 1$ in the simulation volume is given by
\begin{equation}
v [(j)]_{(1)} = \frac{ \frac{z_{(j)}}{j!}{\rm e}^{\beta \mu j} \sum_{l \geq 0} \frac{z_(1)^l {\rm e}^{\beta \mu l}}{l!}}{\left( 1+ \sum_{k>1} \frac{z_{(k)}}{k!}{\rm e}^{\beta \mu k} \right) \sum_{l \geq 0} \frac{z_1^l {\rm e}^{\beta \mu l}}{l!}},
\label{p_j_gcan}
\end{equation}
The numerator in this expression arises from summing Equation \ref{P_eta} for states that contain one cluster of size $j$ and any number of isolated monomers, and the denominator from summing over all possible states containing at most one cluster larger than a single particle. Having simplified the fraction, and using Equation \ref{v[k]}, we are left with 
\begin{equation}
[(j)]_{(1)}= \frac{v^{-1} \frac{z_{(j)}}{j!}{\rm e}^{\beta \mu j} }{\left( 1+ \sum_{k>1} \frac{z_{(k)}}{k!}{\rm e}^{\beta \mu k} \right) } = \frac{[(j)] }{1+ \sum_{k>1} v [(k)]}
\label{[j]sim}
\end{equation}
as the concentration for a system restricted to at most one non-trivial cluster.
We note that this concentration is {\em not} directly comparable to single-target results in the canonical ensemble. As the two are never needed for the same simulations, however, the use of the same notation should not cause confusion.

The relative difference between simulation results and the true behaviour of the model is easy to quantify:
\begin{equation}
\frac{[(j)]-[(j)]_{(1)} }{[(j)]} = \frac{\sum_{k>1} v [(k)]}{1+ \sum_{k>1} v [(k)]}.
\end{equation}
$v [(k)]$ will grow proportionally with the simulation volume. Consequently, the relative errors will initially grow linearly with the simulation volume before plateauing in the limit of $\sum_{k>1} v [(k)] \gg 1$ (when the relative error is approximately unity).
\subsection{Correcting for errors}
\label{Gcanonical-correction}
In this section we show how to extract $[(j)]$ from the measured $[(j)]_{(1)}$. It can be trivially shown that, under our assumptions of ideality, the predicted bulk yield of isolated monomers is the same as the single target yield. Equation \ref{[j]sim} can be rearranged to give $j_{\rm max}-1$ linear simultaneous equations for the remaining $[(j)]$, 
\begin{equation}
\left[
\begin{array}{c c}
1-v [(2)]_{(1)} & -v [(2)]_{(1)} \,\,\,\,\,\,... \\
-v [(3)]_{(1)} & 1-v [(3)]_{(1)} \,... \\
... &\\
\end{array}
\right]
\left[
\begin{array}{c}
[(2)] \\
\left[(3) \right] \\
... 
\end{array}
\right]
=
\left[
\begin{array}{c}
[(2)]_{(1)}\\
\left[(3)\right]_{(1)} \\
... 
\end{array}
\right],
\label{matrix_equation}
\end{equation}
where $j_{\rm max}$ is the largest cluster considered. These equations can then be solved using standard matrix inversion techniques.

In reality, most simulations will not explicitly forbid the presence of multiple clusters of more than one particle (although this can be done, as in Section \ref{example extrapolation GC}). If, however, cluster formation involves a significant free energy barrier, and only the formation of a single cluster is actively biased by the simulation, multiple large clusters will not be observed. In these cases, Equation \ref{[j]sim} can be used, but any rare instances where multiple clusters of more than one particle do occur (for example, two dimers) must not be included in estimating $[(j)]_{(1)}$.
\subsection{Relevance to previous studies}
\label{relevance}
To date, grand canonical techniques (and related semi-grand canonical approaches) have primarily been used to study the formation of micellar structures,\cite{Kusaka1998, Bolhuis1997, Kusaka2001, Pool2005, Kim2001, Floriano99, Panagiotopoulos02, Verde2010, tenWolde1998} as opposed to monodisperse target structures. There is no reason, however, that grand canonical simulations could not be used for monodisperse targets, and the results presented here can be used equally well for both types of assembly.
I
n many cases in the literature, monomer concentrations are assumed to be so low relative to the simulation volume that the probability of finding more than one cluster in a simulation box is neglected in the analysis. The number of monomers in a simulation at a given instant is then taken as a proxy for cluster size.\cite{Kusaka1998, Bolhuis1997, Kusaka2001, Pool2005} There has also been considerable debate on the details of inferring cluster probabilities from simulations in which one monomer is fixed at the centre of the simulation volume, under this extremely dilute assumption.\cite{Kusaka1998, Kusaka1999, Reiss2002}

As the methodology presented here to extract bulk yields is so simple, this extremely dilute assumption seems unnecessary. The risk is that states which are really characteristic of a cluster of size $j$ and another of $j^{\prime}$ are treated as a single cluster of size $j+j^{\prime}$. The frequency of such mis-labelling would tend to increase with simulation volume, resulting in quantitative errors. 

It has been argued that the extremely dilute assumption is valid provided $\sum_k v [(k)] \ll 1$, as the probability of sampling a state with two actual clusters is much smaller than observing either in isolation.\cite{Kusaka1999b} Unfortunately, this is not necessarily the case. For example, consider systems in which $[({j+1})] \ll [(j)]$. In these cases the probability of a volume containing a cluster of size $j$ and an additional isolated particle may be large compared to the probability of observing a genuine cluster of size $j+1$, even if $[(j)]$ and $[(1)]$ are small. There are two cases when this is particularly likely to be relevant:
\begin{enumerate}
\item $[(2)] \ll [(1)]$ is likely to be true for large assemblies, such as micelles, when many particles are needed to stabilize a cluster.
\item $[({j_{0}+1})] \ll [({j_{0}})]$ will be true for monodisperse assemblies where the assembly product has a size of $j_{0}$.
\end{enumerate}
Both of these conditions hold in the system analysed as an example in Section \ref{example extrapolation GC}.

As with the corrections for small systems in the canonical ensemble, applying this methodology will not necessarily give data that seem to match experiments more closely. Rather, these corrections allow data from simulations to be sensibly compared to experiments or theories based on bulk properties. Although in many cases the effects might be quantitative rather than qualitative, given the low computational cost of the correction scheme it would seem sensible to apply it.
\subsection{Implementation issues of the correction scheme}
Are there any drawbacks to using the extrapolation method outlined here? Firstly, it relies upon the assumption that separate clusters can be described as behaving approximately ideally. Such a problem will always arise when bulk physics is extracted from a single self-assembling cluster. Furthermore, as discussed in Section \ref{example extrapolation GC}, this methodology allows the assumption of ideality to be checked.

From the perspective of practical implementation, it is necessary to have an algorithm that evaluates the cluster distribution in a configuration, which is potentially computationally costly. By contrast, if the number of particles in the system is simply taken as a proxy for cluster size, no such calculation is required. To reduce this cost, the clustering could be sampled only every $t \gg 1$ steps of the simulation. If $t$ is similar to the number of steps over which energy correlations within the system are lost, such a reduction of sampling frequency will have a limited effect on simulation accuracy.

\subsection{Example extrapolation}
\label{example extrapolation GC}
\begin{figure}
\begin{center}
\includegraphics[width=6.5cm]{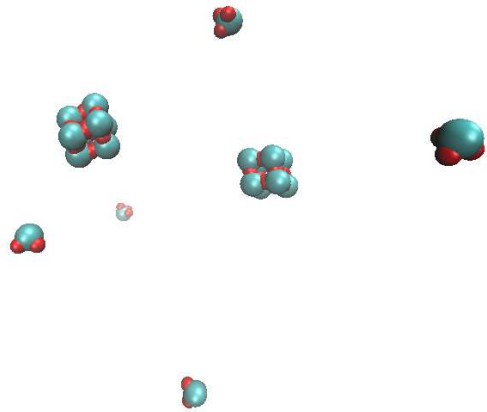}
\caption{\footnotesize{Snapshot from a simulation of the patchy particle model of Wilber {\it et al.}\cite{Wilber09b} The depicted state contains two fully formed cubes (eight-particle clusters) and five isolated monomers.}}
\label{patchy_pic}
\end{center}
\end{figure}
As an illustration of inferring bulk yields from small grand canonical simulations, we consider the patchy-particle model of Wilber {\it et al.}\cite{Wilber09b} As with the DNA simulations in Sections \ref{Example extrapolation C} and \ref{example localized}, we are simply using this model as a typical self-assembling system with which to demonstrate the application of the methodology outlined in the previous sections. 
We consider particles with a patch number and orientation that favours the formation of cubic octamers, as illustrated in Figure \ref{patchy_pic}. The parameters of the model and conditions at which simulations were performed are given in the supplementary material:\cite{supplementary} the specific values were chosen to give a reasonable yield of cubes in a fairly dilute system. 

Initially, simulations were performed in which umbrella sampling was used to accelerate the sampling of a single large cluster, and
states of the system with more than one cluster of multiple particles were explicitly forbidden. More details are provided in the supplementary material.\cite{supplementary} This approach allowed the extrapolation methodology of Section \ref{Gcanonical-correction} to be directly applied. The yields of various cluster sizes, and the corrections to account for multiple large clusters, are given in Table \ref{grand canonical yields}. In this case, all concentrations (except that of isolated monomers) are seen to increase by around 50\,\% due to the extrapolation. An increased concentration from extrapolation is as would be expected: by not sampling states with multiple large clusters, we measure a reduced concentration relative to the true result. 
\begin{table*}
\begin{center}
\begin{tabular}{c | c c c| c c }
size $j$ & $v[(j)]_{(1)}^{\rm sim}$ & $v[(j)]^{\rm pred} $ & $v[(j)]^{\rm sim}_{(\infty)} $ & $v[(j)]_{(2)}^{\rm sim}$ & $v[(j)]_{(2)}^{\rm pred}$ \\ 
\hline
1 & 3.2308(2) & 3.2308(2) & 3.2312(86) &3.2293(2) & 3.2308(2)\\
2 & $1.460(8) \times 10^{-2}$&$2.200(2)\times 10^{-2}$ & $2.192(7)\times 10^{-2}$ & $2.029(9)\times 10^{-2}$ & $2.027(5)\times 10^{-2}$\\
3 &$2.188(12) \times 10^{-4}$ & $3.297(8)\times 10^{-4}$ &$ 3.273(19)\times 10^{-4}$& $3.044(21)\times 10^{-4}$ & $3.038(9)\times 10^{-4}$ \\
4 & $1.032(7) \times 10^{-4}$ & $1.556(9)\times 10^{-4}$ &$1.591(97) \times 10^{-4}$ & $1.431(9)\times 10^{-4}$ & $1.433(8)\times 10^{-4}$\\
5 &$5.239(43) \times 10^{-6}$ & $7.894(52)\times 10^{-6}$ & $7.89(62)\times 10^{-6}$& $7.298(57)\times 10^{-6} $ &$7.273(52)\times 10^{-6} $ \\
6 &$1.603(16) \times 10^{-5}$ &$2.416(21)\times 10^{-5}$ & $2.01(35)\times 10^{-5}$ & $2.249(19)\times 10^{-5}$ &$2.226(20)\times 10^{-5}$ \\
7 &$7.106(53) \times 10^{-5}$ & $1.071(8)\times 10^{-4}$ & $7.320(53)\times 10^{-5}$ &$9.94(11)\times 10^{-4}$ &$9.865(68)\times 10^{-4}$ \\
8 &0.3212(35) & 0.4843(78) &$ 0.3244(41)$ &0.4371(75) & 0.4461(63)) \\
9 & $5.360(80) \times 10^{-5} $&$8.08(14) \times 10^{-5}$ &$5.22(10) \times 10^{-5}$ &$7.31(19) \times 10^{-5}$ &$7.44(12) \times 10^{-5}$ \\
\hline
\end{tabular}
\caption{\footnotesize Yields of various clusters of size $j$ from simulations of a patchy particle model.\cite{Wilber09b} $v[(j)]^{\rm sim}_{(1)}$ is the yield from simulations in which only a single cluster of more than one particle could form. $v[(j)]^{\rm pred}$ is the extrapolation of that result to the limit of an arbitrary number of clusters, performed using Equation \ref{[j]sim}. $v[(j)]^{\rm sim}_{(\infty)}$ is the yield in simulations in which any number of clusters was permitted to form, but biasing was only applied to the largest cluster. $v[(j)]^{\rm sim}_{(2)}$ is the yield from simulations which also accurately sampled states containing two clusters of more than one particle -- this should be compared to the prediction $v[(j)]^{\rm pred}_{(2)}$ obtained by solving Equation \ref{[N_j]sim_2} given $v[(j)]_{(1)}^{\rm sim}$. \label{grand canonical yields}}
\end{center}
\end{table*}

Also shown in Table \ref{grand canonical yields} are the results of simulations in which multiple large clusters were not explicitly forbidden, but which still only used the largest cluster to bias the ensemble and accelerate sampling. In this case, the yield of smaller multiple-particle clusters (with fewer than six particles) is seen to agree well with the extrapolation. Larger clusters, however, do not match the extrapolation and clusters of eight or nine particles have a yield consistent with the one-cluster simulations. This result indicates that these simulations failed to sample states with multiple clusters of this size, due to the large free-energy barrier associated with formation. Accelerating sampling using only the size of the largest cluster is therefore a poor way to equilibrate such a system. The technical difficulty of biasing the formation of an arbitrarily large number of clusters makes the extrapolation procedure a useful alternative.

The accuracy of the extrapolation scheme can be validated by considering the formation of two large clusters. If up to two clusters of more than one particle are sampled, the average number of clusters of size $j$ observed in the simulation is
\begin{equation}
v [{(j)}]_{(1)}=\frac{\frac{z_{(j)}}{j!}{\rm e}^{\beta \mu j}\left(1 +\dSum_{k>1,\,k \neq j} \frac{z_{(k)}}{k!}{\rm e}^{\beta \mu k} + 2\frac{z_{(j)}}{2j!}{\rm e}^{\beta \mu j} \right)}{{1+\dSum_{k\neq l>1} \frac{z_{(k)} z_{(l)}}{2 k! l!}{\rm e}^{\beta \mu (k + l)}} + \dSum_{k>1} \frac{z_{(k)}^2}{2 (k!)^2}{\rm e}^{2 \beta \mu k} }.
\label{gcan 2 clusters}
\end{equation}
This expression follows from considering the contribution to the partition function of all states with two or fewer large clusters. The sum over $k\neq l>1$ is a sum over both $k>1$ and $l>1$, with terms when $k=l$ absent. Note that factors of $\frac{1}{2}$ arise to avoid double counting during sums, and when more than one cluster of the same size is present due to indistinguishability. As in Equation \ref{p_j_gcan}, the contribution of monomer partition functions cancels, and is not included. 

Equation \ref{gcan 2 clusters} simplifies to
\begin{equation}
[(j)]_{(2)} = \frac{[(j)] \left(1+ \sum_{k>1} v [(k)] \right)}{1+ \sum_{k>1} v [(k)] + \frac{1}{2}\left( \sum_{k>1} v [(k)] \right)^2}.
\label{[N_j]sim_2}
\end{equation} 
Further simulations under identical conditions to the original single-cluster simulations were performed. In this case, up to two clusters of more than one particle were allowed, and umbrella sampling was used to bias the size of the two largest clusters. Further details are provided in the supplementary material,\cite{supplementary} and the results are shown in Table \ref{grand canonical yields}, along with the yield predicted by Equation \ref{[N_j]sim_2} using the $v[(j)]^{\rm sim}_{(1)}$ found in the single-cluster simulations. As is evident from Table \ref{grand canonical yields}, the extrapolation method agrees extremely well with the explicit two-target simulations. This strongly suggests that the extrapolation to bulk under these conditions is reliable.

It is possible to detect some very small non-ideal effects. Specifically, for a truly ideal system, $v[(1)] =v[(1)]_{(1)}^{\rm sim} = 3.2372$ for the conditions used here, as described in the supplementary material.\cite{supplementary} Table \ref{grand canonical yields} shows that $v[(1)]_{(1)}^{\rm sim} $ is smaller than this, presumably due to excluded volume effects. This is consistent with the fact that allowing a second large cluster, and thereby increasing excluded volume, suppresses the yield of monomers further ($v[1]_{(2)}^{\rm sim} < v[(1)]_{(1)}^{\rm sim} $). These non-ideal effects, however, are very small (the concentration of monomers is reduced by less than 0.25\,\% relative to the ideal limit in a simulation which allows up to two large clusters). Furthermore, as higher numbers of large clusters in a small volume contribute a limited amount to the partition function, this difference will likely remain small.

In this simulation volume it would be clearly inappropriate to make the approximation that all particles are part of the same cluster, as $\sum_k [(k)] >1$. One could imagine, however, reducing the volume by $\sim$\,$100$, in which case $\sum_k [(k)] \ll 1$ (in fact, a volume this small would probably lead to percolating clusters, but the point it illustrates is generally valid). Even in this limit, however, the 
probability of observing two monomers ($\sim$\,$10^{-3}$) would be larger than observing a genuine dimer ($\sim10^{-4}$), and the probability of observing an 8-particle cluster and a monomer ($\sim$\,$10^{-4}$) would be significantly larger than a 9-particle cluster ($\sim$\,$10^{-7}$). This illustrates that $\sum_k [(k)] \ll 1$ is not enough to justify quantitative yields being inferred from the very dilute approximation, as highlighted in Section \ref{relevance}.

\subsection{Multi-species clusters}
At this stage, only simulations involving one type of particle have been considered. The results in this section are easily extended to simulations of two or more species, each maintained by their own chemical potential. If only one cluster of more than one particle is sampled, the analogous result to Equation \ref{[j]sim} for two species is
\begin{equation}
P(j,k) = \frac{ \frac{z_{(j,k)}}{j!k!}{\rm e}^{\beta (\mu_1 j + \mu_2 k) } }{ 1+ \sum_{p+q>1} \frac{z_{(p,q)}}{p!q!}{\rm e}^{\beta (\mu_1 p+ \mu_2 q) }},
\end{equation}
where $P(j,k)$ is the probability of observing a cluster consisting of $j$ particles of type $1$ and $k$ particles of type $2$. Thus the simulation concentration is
\begin{equation}
[{(j,k)}]_{(1)} = \frac{[({j,k})]}{1 + \sum_{p+q>1} v[({p,q})]}.
\end{equation}
The result is an obvious generalization of the single-species case. The approximations can also be checked by sampling the formation of two clusters, with a yield that is analogous to Equation \ref{[N_j]sim_2}
\begin{equation}
[(j,k)]_{(2)} = \frac{[(j,k)] \left(1+ \dSum_{p+q>1} v [(p,q)] \right)}{1+ \dSum_{p+q>1} v [(p,q)] + \frac{1}{2}\left( \dSum_{p+q>1} v [(p,q)] \right)^2}.
\label{[N_j]sim_2}
\end{equation} 

\section{Inference of properties other than yields}
\label{other properties}
This paper has been hitherto devoted to inferring cluster yields of bulk systems from those found in small simulations. Other properties, such as the average potential energy of the system, or the frequency of a certain type of interaction, may also be of interest. Let us assume that we wish to calculate the thermodynamic average of a quantity $\tilde A$ in bulk, where the tilde indicates that the quantity is normalized per particle in the system. Under the assumptions of the formalism presented here, in which separate clusters do not interact, the internal properties of a given cluster $\{i\}$ are identical in a small system and in the bulk limit. The relative proportions of different clusters do change, however. To calculate the average of $\tilde A$ in the bulk limit, therefore, we must 
measure the average for each cluster type in a small simulation, $\tilde A_{\{i\}}$, then perform a weighted average using the bulk cluster yields inferred via the methods presented in this article.
\begin{equation}
\tilde A = \frac{\sum_{\{i\}} \tilde A_{\{i\}} [\{i\}] i_{\rm tot} }{\sum_{\{i\}} [\{i\}] i_{\rm tot}},
\end{equation} 
where $i_{\rm tot}=\sum_j i_j$ is the total number of particles in a cluster.

\section{Discussion}
In this paper we have extended the methodology of Reference \onlinecite{Ouldridge_bulk_2010} to deal with the inference of bulk properties from small simulations of self-assembly involving multiple particle species and systems in the grand canonical ensemble. In general, bulk systems are directly comparable to experimental studies, but it is often only feasible to simulate assembly of a single target. The methods presented here can be viewed as a process for inferring standard free energies of formation for self-assembling systems from small simulations, and checking the accuracy of the ideal assumptions underlying that inference. 

For simulations of a single self-assembling cluster in the canonical ensemble, large deviations from the bulk yield that would be found for the same model are observed due to neglected concentration fluctuations. These errors can be corrected using the methodology presented here, under the assumption that separate clusters behave ideally. If the formation of two or more clusters can be studied, the accuracy of this assumption can be checked by examining the convergence on the large system limit. As with clusters of one species of particle,\cite{Ouldridge_bulk_2010} convergence on the bulk limit as system size increases can be very slow, particularly if one cluster-type dominates the ensemble.

As a consequence, if quantitative data is to be extracted from canonical simulations of a single cluster, this methodology (or something equivalent) should be applied. A summary of the necessary steps for extrapolating results to the bulk limit from a single-target canonical simulation is:
\begin{itemize}
\item Perform a single-target simulation in a volume $v$, measuring the cluster frequency $v[\{i\}]_{(1)}$.
\item Obtain $\psi_{\{i\}}$ by fitting the measured $v[\{i\}]_{(1)}$ using Equation \ref{D_equals_1_equation}. 
\item Solve for the bulk concentration $[\{i\}]$ using $v$ and $\psi_{\{i\}}$ in Equation \ref{thermo_multi_particle_2} whilst fixing the total concentrations using Equation \ref{conc_yield}. 
\end{itemize}
Furthermore, if a canonical simulation is performed in which several clusters can form ($d$ possible clusters in a volume $dv$), the relevance of statistical finite size effects can be assessed in the following manner:
\begin{itemize}
\item First, assume the observed cluster distributions are reflective of bulk concentrations $[{\{i\}}]$.
\item Use these $[{\{i\}}]$ to estimate the ratios $\psi_{\{i\}}$, using Equations \ref{thermo_multi_particle_2} and \ref{conc_yield}.
\item Follow the convergence of cluster yields on the bulk limit using Equation \ref{D_greater_1_equation}. If the yield at size $d$ is consistent with the initial results, then the consequences of simulating a finite number of clusters are likely to be negligible. If, however, the results are not self-consistent, then finite size effects remain significant at a system size $d$.
\end{itemize}

Extending the analysis of Reference \onlinecite{Ouldridge_bulk_2010} to many constituent species also allows us to treat the special case in which one reactant is localized in space. We find that the extrapolation procedure is unchanged if only one species, that contributes at most one particle to any cluster, is localized. This is particularly relevant to several studies involving binding of DNA to strands that are localized on a surface, such as References \onlinecite{Tito2010, Allen2011, Hoefert2011, Schmitt2011}. None of these groups included a finite size correction, and hence the quantitative results are not directly applicable to bulk systems. We note that the density of adsorbed strands on DNA microarray surfaces, a common case in which localization is relevant, is usually high enough to cause interactions that invalidate the assumptions in this work. Such physics, however, was not explored in References \onlinecite{Tito2010, Allen2011, Hoefert2011, Schmitt2011} as only a single target was considered. These investigations can therefore only be sensibly compared with the low-density limit, for which the finite-size corrections presented in this work are appropriate.

We have also considered simulations in the grand canonical ensemble. This technique naturally incorporates the concentration fluctuations that were absent in canonical systems. Bulk yields, however, can only be obtained if multiple large clusters can form in the simulation volume, a process which may be difficult to sample. We have shown how to use the data collected for the assembly of one cluster to infer the bulk yield under the usual ideal approximations, and how to check those approximations if two clusters can be simulated. The procedure for extrapolating to bulk when only one large cluster can be simulated is:
\begin{itemize}
\item Perform a simulation in a volume $v$ in which only a single target cluster is sampled. Measure the cluster frequency $v[\{i\}]_{(1)}$, ignoring any states that contain multiple large clusters.
\item The bulk concentration of monomers under these conditions is given by the concentration in the single-target simulation. To obtain all other concentrations, use the measured $v[\{i\}]_{(1)}$ to construct the matrix Equation \ref{matrix_equation}, and solve for $[\{i\}]$ by inverting it. 
\item If desired, $\psi_{\{i\}}$ can be extracted from Equation \ref{thermo_multi_particle_2} using $[\{i\}]$, allowing the bulk yield to be estimated at other concentrations.
\end{itemize}

An alternative technique that has been used in the past for grand canonical simulations is to assume that any state in which $j$ particles are in the simulation volume corresponds to a $j$-particle cluster. This assumption of extreme dilution simplifies the simulation (there is no need to define or measure clusters within the simulation), but it can be applied over a much smaller range of conditions. It has been claimed that it is valid provided the total number of particles in a simulation volume is small: we have shown, however, that even in this limit large quantitative errors can exist. By contrast, the technique demonstrated here is valid whenever separate clusters behave in an approximately ideal fashion, and allows the accuracy of the extrapolation to be quantitatively assessed.

In addition to the theoretical analysis presented here, we have shown examples of typical self-assembling systems for which the bulk yield can be accurately inferred. We have demonstrated that the corrections are practically implementable, and that the accuracy of the underlying assumptions can be reasonably assessed. We have also outlined a procedure for inferring bulk properties other than the yields of clusters, such as the average potential energy, using values measured in single-target simulations.

In some cases the cluster yields obtained using the methodology presented here may be described as only quantitatively, rather than qualitatively, different from the single-target data. Nonetheless, if comparisons of models with bulk experiment are to be made, then it is sensible to apply these corrections, as failure to do so is analogous to reporting results obtained with a faulty algorithm that causes a quantitative error. Although mesoscale models are never going to give precise descriptions of all the properties of a system, many have recently been used to provide quantitative comparisons of yields with experiment,\cite{Ouldridge2009, Ouldridge_tweezers_2010, Ouldridge2011, Sambriski2008, Sambriski2009, Prytkova2010, Freeman2011, Tito2010, Allen2011, Hoefert2011, Schmitt2011, Romano2012} and hence should consider the corrections presented here. Other authors have not compared to experiment, but have related equilibrium thermodynamics obtained from single-target simulations to bulk simulations of the same model.\cite{Wilber2007,Wilber09b} The effects discussed in this work are relevant to such a self-consistent comparison. As computational power increases, the simulation of self-assembly for more detailed models will become possible: to compare these approaches with experiment, whether to make predictions or validate and parameterize force fields, finite-size corrections may be relevant. Finally, as the inference of the bulk yields generally requires much less effort than performing the simulations themselves, it seems sensible to do it in all cases.
\acknowledgements
The author would like to acknowledge helpful input from Aleks Reinhardt, Ard Louis and Jonathan Doye, and funding from University College, Oxford.

\appendix{}
\section{Mesoscale models used in the examples}
\subsection{DNA}
\label{DNA}
The examples of DNA self-assembly  presented in the main paper involve the coarse-grained model of Reference \cite{Ouldridge_tweezers_2010}, using its most recent parameterization in Reference \cite{Ouldridge_thesis}. In short, the model treats DNA as a string of rigid nucleotides which interact through physically motivated pairwise contributions to the energy. The rigid nucleotides contain interaction sites to represent the sugar-phosphate backbone and the base. Of particular importance for our purposes are the hydrogen-bonding interactions, which allow base pairs to form between nucleotides. Bases come in four types: adenine (A), guanine (G), cytosine (C) and thymine (T). In this model, AT and GC can form complementary base pairs through hydrogen-bonding interactions. This base pairing leads to the formation of double-helical bound states for two strands with complementary sequences.

The short-ranged nature of the interactions in the model means that there is a clear distinction between bound states of two strands, with a substantial energy of interaction, and unbound states, with no interaction energy. In all cases we consider two strands to be bound if there is at least one hydrogen-bonding interaction with an energy more negative than $-0.60$\,kcal\,mol$^{-1}$, about $1/7$ of a typical hydrogen-bonding interaction. The results presented are not sensitive to the precise value of this cutoff, as the cooperativity of helix formation means that the overwhelming majority of bound pairs have well-formed duplexes.

\subsection{Patchy particles}
\label{patchy details}
The demonstration of self-assembly involving cubic octamers in Section III.F of the main paper used the patchy particle model of Reference \cite{Wilber09b}. In this model, particles have a number of `patches' distributed over a spherical surface with a symmetry that determines the structure of stable clusters. Two particles interact through short-ranged repulsion and medium-range attraction, the latter being modulated by terms related to the angular and torsional alignment of the best-aligned pair of patches on the particles. 

\begin{figure}
\begin{center}
\includegraphics[width=4cm]{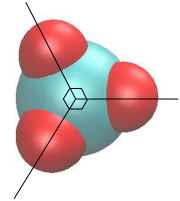}
\caption{\footnotesize{ A patchy particle which tends to form cubic octamers. The centre of the large sphere represents the particle's centre of mass. The smaller spheres representative interactive patches. Note that the smaller spheres are simply illustrative of patch location: the patches have no actual volume.}}
\label{patchy figure}
\end{center}
\end{figure}

Particles with three patches as shown in Figure \ref{patchy figure} have a tendency to form cubic clusters: the inclusion of angular and torsional modulation of interactions disfavours alternatives such as large aggregates \cite{Wilber09b}. In this work we use  the parameters given in Table \ref{gpatchy parameters} as convenient choices within the range of values considered in Reference \cite{Wilber09b}. Note, however, that we use much lower concentrations (by a factor of $\sim$\,$250$) than typically studied in Reference \cite{Wilber09b}. This difference ensures that the ideal assumptions required for extrapolation remain valid in our simulations. At higher concentrations, differences between single-target and bulk systems persist but cannot be accurately treated within the framework of the main article. As pointed out by Wilber {\it et al.}, however, the concentrations used in Reference \cite{Wilber09b} are artificially high in order to accelerate the kinetics of assembly.

\begin{table}
\begin{center}
\begin{tabular}{c | c }
Parameter name & Value \\ \hline
$\epsilon$ & 1 \\
$\sigma_{\rm LJ}$ & 1 \\
$\sigma^2_{\rm ang}$ & 0.2 \\
$\sigma^2_{\rm tor}$ & 0.4 \\
\end{tabular}
\caption{\footnotesize Specific parameters of the patchy particle model of Reference \cite{Wilber09b} used in this study. $\epsilon$ sets the energy scale of the interaction between particles, $\sigma_{\rm LJ}$ the range and $\sigma^2_{\rm ang}$ and $\sigma^2_{\rm tor}$  the width and torsional tolerance of the patches on the particles. \label{gpatchy parameters}}
\end{center}
\end{table}

During simulations, particles with an interaction energy $E_{ij} < -0.1 \epsilon$, with $\epsilon$ being the energy scale of the interaction, were considered to be part of the same cluster. The results obtained are not sensitive to small changes in this value.

\section{Simulation methods}
\subsection{Metropolis Monte Carlo}
The Metropolis Monte Carlo algorithm (MC) \cite{metropolis1953} is a widely used method for calculating the thermodynamic properties of a system. From a given initial state, a trial move to another state is selected and accepted with a probability that ensures the algorithm samples from the Boltzmann distribution. The simplest MC algorithms attempt state changes that involve altering a single object within the system -- for instance the position or orientation of a particle in a molecular simulation.

\subsection{Virtual Move Monte Carlo}
Simple MC algorithms can face difficulty in reaching equilibrium if the collective motion of strongly-interacting particles is required, as such collective motion is slow when only single-particle moves are attempted. Algorithms which attempt to move clusters of particles can overcome this difficulty: one example is the `Virtual Move Monte Carlo' (VMMC) algorithm \cite{Whitelam2007}, which dynamically generates clusters of particles based on energy changes from trial moves. When simulating DNA, we use the variant of VMMC in the appendix of Reference \cite{Whitelam2009}.

\subsection{Umbrella sampling}
Even with VMMC, self-assembly processes can be slow to equilibrate due to high free-energy barriers during formation. Umbrella sampling \cite{Torrie1977}, which involves imposing an artificial biasing weight  $W({\bf r}^N)$ on a system with degrees of freedom ${\bf r}^N$, can be used to reduce the effective height of the barrier. A lower barrier means transitions occur more quickly, and equilibration is accelerated. The thermodynamic expectation of any variable $A$ follows from the biased sample obtained as
\begin{equation}
\langle A \rangle = \frac{\langle A({\bf r}^N)/W({\bf r}^N) \rangle_W }{\langle 1/W({\bf r}^N) \rangle_W}.
\end{equation}
Here $\langle \rangle_W$ indicates the expectation found by sampling from the biased distribution \linebreak $W({\bf r}^N) \exp(-\beta U({\bf r}^N))$, with $U({\bf r}^N)$ being the internal energy.

\section{Simulation details}
\subsection{DNA simulation}
\label{DNA_simulation}
The DNA model was simulated using the VMMC algorithm, with initial trial moves being either:
\begin{itemize}
\item Rotation of a nucleotide about its backbone site, with the axis chosen from a uniform random distribution and the angle from a normal distribution with mean of zero and a standard deviation of 0.2 radians.
\item Translation of a nucleotide with the direction chosen from a uniform random distribution and the distance from a normal distribution with mean of zero  and a standard deviation of 1.7\,\AA.
\end{itemize}

\subsubsection{DNA trimer formation}
\label{DNA_simulation_trimer}
The simulation of DNA trimers involved three distinct strands:
\begin{enumerate}
\item $5^\prime$-GACGACTTAAGGAG-$3^\prime$
\item $5^\prime$-CTCCTTTTCGACCG-$3^\prime$ 
\item $5^\prime$-CGGTCGTTGTCGTC-$3^\prime$
\end{enumerate}
Here, the sequence specifies the base of each nucleotide: adenine (A), guanine (G), thymine (T) or cytosine (C). The $3^\prime$ and $5^\prime$ symbols indicate strand directionality. The Watson-Crick rules of complementarity, which are incorporated into the model, dictate that strong bonds can only form between AT and CG pairs \cite{Saenger1984}. As a result, the three strands tend to form three-armed junctions, as each strand has two 6-base sections that are complementary to 6-base sections on different strands.

All simulations were performed at 307.7\,K, in a simulation volume of $1.669 \times 10^{-23}$\,m$^3$ per trimer. 20 and 40 simulations were performed for the assembly of one and two trimers respectively, using $4\times 10^{10}$ attempted moves of the VMMC algorithm each. Umbrella sampling was used to enhance equilibration. For the single-target simulations, the umbrella potential was given by
\begin{equation}
W = D_1(s_1)E_1(p)F(n_b - n_c).
\end{equation}
Here $s_1$ represents the size of the largest cluster in the system (in terms of number of strands), $p$ the smallest non-zero number of base pairs between any two strands, $n_c$ the number of clusters and $n_b$ the number of pairs of interacting strands. The functional form of $D_1$ is 
\begin{equation}
D_1 = 
\left\{
\begin{array}{c c}
10 & {\rm if }\,\,s_1=1\\
1 & {\rm if }\,\,s_1=2\\
25 & {\rm if }\,\,s_1=3\\
\end{array}
\right\}.
\end{equation}
F is given by $F(n_b - n_c) = 0.04^{(n_b-n_c)}$, and $E_1$ is given in Table \ref{E_table}.
For two-target simulations, the umbrella potential was also a function of the size of the second largest cluster in the system, $s_2$: $W = D_2(s_1,s_2)E_2(p)F(n_b - n_c)$. $F$ has the same definition as for the single-target case, $E_2$ is defined in Table \ref{E_table} and $D_2$ in Table \ref{D_table}.

\begin{table}
\begin{center}
\begin{tabular}{c |c c c c c c c c  }
&  \multicolumn{8} {c} {$p$} \\
 & 0 & 1&  2& 3& 4& 5& 6& $\geq7$ \\ \hline
$E_1(p)$& 1&2000&400&50&10&5&3&1\\
$E_2(p)$ & 1 &1000&150&30&5&2&3&1 \\
\end{tabular}
\caption{\footnotesize The functions $E_1(p)$ and $E_2(p)$ used in the umbrella potential for simulations of DNA trimer formation. \label{E_table}}
\end{center}
\end{table}

\begin{table}
\begin{center}
\begin{tabular}{c c | c c c c c c}
\multicolumn{2}{c|}{$D_2(s_1,s_2)$}& \multicolumn{6} {c} {$s_1$} \\
& &  1&  2& 3& 4& 5& 6 \\ \hline 
& 0 &   &  &  &  &  & $10^6$\\
$s_2$ & 1  & 10 & 0.8 & 25 & $10^3$ & $3 \times 10^4$ \\
& 2 & &  0.2 & 15 & $10^3$ \\
& 3 & & & $5\times 10^3$\\
\end{tabular}
\caption{\footnotesize The function $D_2(s_1,s_2)$ used in the umbrella potential for two-target simulations of DNA trimer formation. Values of $s_1,s_2$ with no entry are impossible. \label{D_table}}
\end{center}
\end{table}

In order to perform the extrapolation, $\psi_{\{i\}}$ must be fitted. This fitting was  performed by minimizing
\begin{equation}
 \sum_{\{i\}} \left(\log  \left( \frac{ v[\{i\}]_{(1)}^{\rm sim}}{v[\{i\}]_{(1)}^{\rm fit}} \right) \right)^2,
\end{equation}
with $v[\{i\}]^{\rm sim}$ being the measured average number of clusters of type $\{i\}$ in a single-target simulation, and $v[\{i\}]^{\rm fit}$ being the estimate obtained from Equation 11 of the main text for a given set of $\psi_{\{i\}}$. The minimization was performed using the  Matlab `fminsearch' function. 

 Due to the biasing umbrella potential, different simulations with the same number of VMMC steps had different overall statistical weight. The data reported in Table I of the article therefore involve weighted estimates of the mean and standard error, using the ``ratio estimate" of Cochran \cite{Cochran1959}. Extrapolation was performed individually for each single-trimer simulation, and averaged using the same weighting factors.

\subsubsection{DNA duplex with one strand localized}
\label{DNA_simulation_localized}
The simulation of DNA with one localized strand used two sequences:
\begin{enumerate}
\item $5^\prime$-TTTAGCTCA-$3^\prime$
\item $5^\prime$-TGAGCT-$3^\prime$ 
\end{enumerate}
Single-target simulations were performed in a cubic periodic cell of volume $1.669 \times 10^{-23}$\,m$^3$ and at a temperature of 300\,K. To model a surface to which strands might be attached, an infinite energy penalty was imposed upon any backbone sites that entered the region $|z| < 8.5$\,\AA. The longer of the two sequences was attached (via the backbone site at the $5^\prime$ end) to the point (0,\,\AA\,0\,\AA,\,8.5\,\AA)  by a harmonic spring with a spring constant 0.571\,Nm$^{-1}$ and equilibrium length 8.5\,\AA. For two-target simulations, a cell of twice the volume was used and the second strand of type 1 was attached at (161\,\AA,\,161\,\AA,\,8.5\,\AA), far enough away from the first to avoid any interaction. 

\begin{table}
\begin{center}
\begin{tabular}{c c |c c c c c c c c c c c }
\multicolumn{2}{c|}{$G_1(t,c)$}& \multicolumn{11} {c} {$t$} \\
& & 0 & 1&  2& 3& 4& 5& 6& 7& 8& 9& $\geq 10$  \\ \hline 
& 0 & 3 & 300&100&20&3&1&1&1&1&1&0 \\
& 1 & & $3.6\times10^4$&100&20&3&1&1&1&1&1&0\\
& 2 & & & 1800&20&3&1&1&1&1&1 &0\\
$c$ & 3 & & & & 140&3&1&1&1&1&1&0\\
& 4 & & & & &  10&1&1&1&1&1&0\\
& 5 & & & & & & 3&1&1&1&1&0\\
& 6 & & & & & & & 1&1&1&1&0\\
\end{tabular}
\caption{\footnotesize The function $G_1(t,c)$ used in the umbrella potential for simulations of DNA duplex formation involving a tethered particle. Values of $c>t$ are impossible. \label{G_table}}
\end{center}
\end{table}

\begin{table}
\begin{center}
\begin{tabular}{c c |c c c c c c c c c c c }
\multicolumn{2}{c|}{$G_2(t,c)$}& \multicolumn{11} {c} {$t$} \\
& & 0 & 1&  2& 3& 4& 5& 6& 7& 8& 9& $\geq 10$  \\ \hline 
& 0 & 3 & 3000&1000&200&30&3&1&1&1&1&0 \\
& 1 & & $3 \times10^4$&1000&200&30&3&1&1&1&1&0\\
& 2 & & & 2000&200&30&3&1&1&1&1 &0\\
$c$ & 3 & & & & 200&30&3&1&1&1&1&0\\
& 4 & & & & &  15&3&1&1&1&1&0\\
& 5 & & & & & & 3&1&1&1&1&0\\
& 6 & & & & & & & 1&1&1&1&0\\x
\end{tabular}
\caption{\footnotesize The function $G_2(t,c)$ used in the umbrella potential for simulations of two-target DNA duplex formation involving a tethered particle. Values of $c>t$ are impossible. \label{G_2_table}}
\end{center}
\end{table}

We performed 10 single-target and 20 two-target simulations, each consisting of $4\times 10^{10}$ attempted VMMC steps. Umbrella sampling was used to accelerate equilibration. In the single-target simulation, the bias used was $W = G_1(t,c)$, with $t$ being the total number of bonds formed between the free strand and the tethered stand, and $c$ being the number of these that are intended to form in the final (fully-aligned) structure (mis-aligned bonds can form, but only contribute to $t$, not $c$). The functional form of $G$ is given in Table \ref{G_table}. In the two-target simulation, $W = G_2(t_1,c_1) G_2(t_2,c_2)$ was used, with $t_i$ being the total number of bonds between tethered strand $i$ and {\em either} of the free strands. The functional form of $G_2$ is given in Table \ref{G_2_table}.  The distribution of clusters was recorded at each step. As with the DNA trimers, weighted estimates of the mean and standard error of inferred yields are reported in the text.

\subsection{Cubic octamers formed from patchy particles}
\label{patchy simulation}
Patchy particle simulations were performed using a simple MC algorithm, with additional moves for removal and addition of particles to make the ensemble grand canonical \cite{Frenkel2001}. Specifically, the attempted moves were:
\begin{itemize}
\item Rotation of a particle about its centre, with the axis chosen from a uniform random distribution and the angle from a normal distribution with mean of zero and a standard deviation of 0.2 radians.
\item Translation of a particle with the direction chosen from a uniform random distribution and the distance from a normal distribution with mean of zero  and standard deviation of $0.2\sigma_{\rm LJ}$
\item Addition of a particle with a randomly chosen position and orientation.
\item Removal of a randomly chosen particle.
\end{itemize}
Simulations were performed at a reduced temperature of $T=0.08$, in a periodic cubic cell of volume $8000\sigma_{\rm LJ}^3$ and with a chemical potential given by $\frac{v \Omega}{\Lambda^3 \Lambda^3_{\Omega}} \exp(\beta \mu)=3.2372$. Here $\Lambda$ is the de Broglie wavelength of the particles, and $\Omega \Lambda^{-3}_{\Omega}$ is the contribution of the angular and angular momenta degrees of freedom to the partition function of an isolated monomer. Defining $\mu$ in this way renormalizes it so that the particle masses and moments of inertia do not need to be considered. The choice of simulation parameters ensured that both monomers and octamers occurred in the simulation box with reasonable frequency.

Umbrella sampling was used to accelerate equilibration. In the case of simulations which allowed a single cluster of more than one particle, the umbrella bias was given by
\begin{equation}
W = U_1(s_1) V(s_1,e_1) \theta(n_c-2).
\end{equation}
In this equation, $s_1$ represents the number of particles in the largest cluster, $e_1$ the interaction energy between particles in that cluster and $n_c$ the number of clusters containing more than one particle. The functional form of $U_1$ is given in Table \ref{U_1 and U_2}. The functional form of $V$ is
\begin{equation}
V(s_1,e_1) =
\left\{
\begin{array}{c c}
100 & {\rm if} \,\, s_1=8 \,\, {\rm and} \,\, e_1<-8 \\
1 & {\rm otherwise} \\
\end{array}
\right\},
\end{equation}
and the functional form of $\theta$ is
\begin{equation}
\theta(x) =
\left\{
\begin{array}{c c}
1 & {\rm if} \,\, x<0 \\
0 & {\rm otherwise} \\
\end{array}
\right\}.
\end{equation}

For the simulations that allowed an arbitrary number of clusters to form, but only biased the formation of one cluster, the umbrella potential was identical except for the $\theta$ term, which was not included. In the two-target case, $W({\bf r}^N)$  also depended on the size of the second largest cluster $s_2$, and its energy $e_2$.
\begin{equation}
W = U_2(s_1,s_2) V(s_1,e_1) V(s_2,e_2) \theta(n_c-3).
\end{equation}
The functional form of $U_2$ is given in Table \ref{U_1 and U_2}, and $V$ and $\theta$ are defined as before.

\begin{table*}
\begin{center}
\begin{tabular}{c c |c c c c c c c c c c }
\multicolumn{2}{c|}{$U_1(s_1)$}& \multicolumn{10} {c} {$s_1$} \\
& & 0 & 1&  2& 3& 4& 5& 6& 7& 8&$\geq 9$ \\ \hline
& & 3 & 0.5 & 15 &$10^3$ & $2 \times 10^3$ & $5\times 10^4$ & $1.5 \times 10^4$ & $3 \times 10^3$ & 1.5 & 1\\
& & &\\
& & & \\
\multicolumn{2}{c|}{$U_2(s_1,s_2)$}& \multicolumn{10} {c} {$s_1$} \\
& & 0 & 1&  2& 3& 4& 5& 6& 7& 8& $\geq 9$ \\ \hline 
& 0 & 2 & 4 & 100 & $10^4$ & $ 4 \times 10^4$ & $8 \times 10^5$ &$10^5$ & $ 2 \times 10^4$ & 25 & 1 \\
& 1 &  &  0.5 & 15 &$10^3$ & $2 \times 10^3$ & $5\times 10^4$ & $1.5 \times 10^4$ & $ 1.5 \times 10^3$ & 1 & 1\\
& 2 &  & & 300 & $10^4$ & $ 2 \times 10^4$ & $5 \times 10^5$ & $1.5 \times 10^5$ & $ 3 \times 10^4$ & 20 & 1\\
& 3 & & & & $10^6$ & $2 \times 10^6$ &  $5 \times 10^7$ & $1.5 \times 10^7$ & $1.5 \times 10^6$ & 500 & 1\\
& 4 & & & &  & $4 \times 10^6$ &  $5 \times 10^7$ & $1.5 \times 10^7$ & $3 \times 10^6$ & 2000 & 1\\
& 5 &  & & & & & $3 \times 10^9$ & $3 \times 10^8$ & $1 \times 10^8$ & $5 \times 10^4$ & 1\\
$s_2 $& 6 & & & & & & &  $3 \times 10^8$ & $4 \times 10^7$ &$10^4$& 1\\
& 7 & & & & & & & & $10^7$ &$3 \times 10^3$ & 1 \\
& 8 & & & & & & & & & 4 & 1  \\
& $\geq 9$ & & & & & & & & & & 1\\ 
\end{tabular}
\caption{\footnotesize The functions $U_1(s_1)$ and $U_1(s_1,s_2)$ used in the umbrella potential for simulations of the cubic octamer formation from patchy particles $s_2>s_1$ is impossible. \label{U_1 and U_2}}
\end{center}
\end{table*}

In the case of single-target simulations, 10 independent runs were performed of $10^{11}$ MC steps each. For two-target simulations, 20 runs of $10^{11}$ MC steps were used. As with the DNA simulations, weighted estimates of the mean and standard error of cluster yields  were calculated by pooling the independent estimates. Matrix inversion was performed using the Matlab `inv' function.

\end{document}